\newcommand{\spectraProcessed}{280~million}
\newcommand{\starProcessed}{9.8~million}
\newcommand{\starPublished}{7\,224\,631}
\newcommand{\meanCompleteness}{77.2\%}
\newcommand{\medianUncertainty}{1.05}

\newcommand{\grvs}{$G_\mathrm{RVS}$}
\newcommand{\obgrvs}{$G_\mathrm{RVS}^{\rm{on-board}}$}
\newcommand{\extgrvs}{$G_\mathrm{RVS}^{\rm{ext}}$}
\newcommand{\intgrvs}{$G_\mathrm{RVS}^{\rm{int}}$}

\newcommand{\Teff}{T_{\rm eff}}
\newcommand{\logg}{\log g}
\newcommand{\FeH}{[Fe/H]}

\newcommand{\tempTeff}{$T_{\rm eff}^{\rm tpl}$}

\newcommand\gaia{\textit{Gaia}}
\newcommand\gdrone{\gaia~DR1}
\newcommand\gdrtwo{\gaia~DR2}
\newcommand\gdrthree{\gaia~DR3}
\newcommand\gdrfour{\gaia~DR4}
\newcommand\RVS{\textit{Radial Velocity Spectrometer}}

\newcommand\kms{\ensuremath{\text{km~s}^{-1}}}
\newcommand\ms{\ensuremath{\text{m~s}^{-1}}}

\documentclass[longauth]{aa}

\usepackage{wrapfig}
\usepackage{longtable}
\usepackage{rotating}
\usepackage{lscape}
\usepackage{amsmath}
\usepackage{adjustbox}
\usepackage{subcaption} 

\usepackage{graphicx}
\usepackage{txfonts}
\usepackage{todonotes}
\usepackage{pdflscape}
\usepackage{rotating}
\usepackage[font=small,labelfont=bf]{caption}

\begin{document} 
\newcommand\vtidy{\vspace{-\baselineskip}}
\newcommand\etc{{\it etc.\ }}
\newcommand\eg{e.g.\ }
\newcommand\ie{i.e.\ }

\title{{\gaia} Data Release 2\\ Properties and validation of the radial velocities}

\author{
D.~Katz                    \inst{\ref{gepi}}    
\and P.~Sartoretti              \inst{\ref{gepi}}
\and M.~Cropper                 \inst{\ref{mssl}}  
\and P.~Panuzzo                 \inst{\ref{gepi}} 
\and G.M.~Seabroke              \inst{\ref{mssl}}
\and Y.~Viala                   \inst{\ref{gepi}}
\and K.~Benson                  \inst{\ref{mssl}}  
\and R.~Blomme                  \inst{\ref{brussels}}
\and G.~Jasniewicz              \inst{\ref{montpellier}}
\and A.~Jean-Antoine            \inst{\ref{cnes}}
\and H.~Huckle                  \inst{\ref{mssl}}  
\and M.~Smith                   \inst{\ref{mssl}}
\and S.~Baker                   \inst{\ref{mssl}} 
\and F.~Crifo                   \inst{\ref{gepi}}  
\and Y.~Damerdji                \inst{\ref{alger},\ref{liege}}
\and M.~David                   \inst{\ref{antwerpen}}
\and C.~Dolding                 \inst{\ref{mssl}}
\and Y.~Fr\'{e}mat              \inst{\ref{brussels}}\relax
\and E.~Gosset                  \inst{\ref{liege},\ref{fnrs}}\relax
\and A.~Guerrier                \inst{\ref{thales}}
\and L.P.~Guy                   \inst{\ref{geneva}}
\and R.~Haigron                 \inst{\ref{gepi}}
\and K.~Jan{\ss}en              \inst{\ref{aip}}
\and O.~Marchal                 \inst{\ref{gepi}}
\and G.~Plum                    \inst{\ref{gepi}}
\and C.~Soubiran                \inst{\ref{bordeaux}}
\and F.~Th\'{e}venin            \inst{\ref{oca}}
\and M.~Ajaj                    \inst{\ref{gepi}}  
\and C.~Allende Prieto          \inst{\ref{mssl},\ref{tenerife},\ref{tenerife2}}
\and C.~Babusiaux               \inst{\ref{gepi},\ref{grenoble}}
\and S.~Boudreault              \inst{\ref{mssl},\ref{maxplanck}}
\and L.~Chemin                  \inst{\ref{bordeaux},\ref{chile}}
\and C.~Delle Luche             \inst{\ref{gepi},\ref{thales}}
\and C.~Fabre                   \inst{\ref{atos}}
\and A.~Gueguen                 \inst{\ref{gepi},\ref{garching}} 
\and N.C.~Hambly                \inst{\ref{edinburgh}}
\and Y.~Lasne                   \inst{\ref{cnes}}
\and F.~Meynadier               \inst{\ref{gepi},\ref{syrte}}
\and F.~Pailler                 \inst{\ref{cnes}}
\and C.~Panem                   \inst{\ref{cnes}}
\and F.~Royer                   \inst{\ref{gepi}}  
\and G.~Tauran                  \inst{\ref{cnes}}
\and C.~Zurbach                 \inst{\ref{montpellier}}\relax
\and T.~Zwitter                 \inst{\ref{ljubljana}}\relax 
\and F.~Arenou                  \inst{\ref{gepi}}
\and D.~Bossini                 \inst{\ref{padova}}
\and A.~Gomez                   \inst{\ref{gepi}} 
\and V.~Lemaitre                \inst{\ref{cnes}}
\and N.~Leclerc                 \inst{\ref{gepi}}
\and T.~Morel                   \inst{\ref{liege}} 
\and U.~Munari                  \inst{\ref{asiago}} 
\and C.~Turon                   \inst{\ref{gepi}}
\and A.~Vallenari               \inst{\ref{padova}}
\and M.~\v{Z}erjal              \inst{\ref{ljubljana},\ref{canberra}}
}     
\institute{
GEPI, Observatoire de Paris, Universit\'e PSL, CNRS, 5 Place Jules Janssen, 92190 Meudon, France
\relax\label{gepi}
\and 
Mullard Space Science Laboratory, University College London, Holmbury St Mary, Dorking, Surrey RH5 6NT, United Kingdom
\relax\label{mssl}
\and 
Royal Observatory of Belgium, Ringlaan 3, 1180 Brussels, Belgium
\relax\label{brussels}
\and 
Laboratoire Univers et Particules de Montpellier, Universit\'{e} Montpellier, CNRS, Place Eug\`{e}ne Bataillon, CC72, 34095 Montpellier Cedex 05, France
\relax\label{montpellier}
\and 
CNES Centre Spatial de Toulouse, 18 avenue Edouard Belin, 31401 Toulouse Cedex 9, France
\relax\label{cnes}
\and 
CRAAG - Centre de Recherche en Astronomie, Astrophysique et G\'{e}ophysique, Route de l'Observatoire Bp 63 Bouzareah 16340, Alger, Alg\'erie
\relax\label{alger}
\and 
Institut d'Astrophysique et de G\'{e}ophysique, Universit\'{e} de Li\`{e}ge, 19c, All\'{e}e du 6 Ao\^{u}t, B-4000 Li\`{e}ge, Belgium
\label{liege}
\and 
Universiteit Antwerpen, Onderzoeksgroep Toegepaste Wiskunde, Middelheimlaan 1, 2020 Antwerpen, Belgium
\relax\label{antwerpen}
\and
F.R.S.-FNRS, Rue d'Egmont 5, 1000 Brussels, Belgium
\label{fnrs}
\and
Thales Services for CNES Centre Spatial de Toulouse, 18 avenue Edouard Belin, 31401 Toulouse Cedex 9, France
\label{thales}
\and
Department of Astronomy, University of Geneva, Chemin d'Ecogia 16, CH-1290 Versoix, Switzerland
\relax\label{geneva}
\and
Leibniz Institute for Astrophysics Potsdam (AIP), An der Sternwarte 16, 14482 Potsdam, Germany
\label{aip}
\and Laboratoire d'astrophysique de Bordeaux, Universit\'{e} de Bordeaux, CNRS, B18N, all{\'e}e Geoffroy Saint-Hilaire, 33615 Pessac, France
\label{bordeaux}
\and
Laboratoire Lagrange, Universit\'{e} Nice Sophia-Antipolis, Observatoire de la C\^{o}te d'Azur, CNRS, CS 34229, F-06304 Nice Cedex, France
\label{oca}
\and 
Instituto de Astrof\'{\i}sica de Canarias, E-38205 La Laguna, Tenerife, Spain
\label{tenerife}
\and
Universidad de La Laguna, Departamento de Astrof\'{\i}sica, E-38206 La Laguna, Tenerife, Spain
\label{tenerife2}
\and
Univ. Grenoble Alpes, CNRS, IPAG, 38000 Grenoble, France
\label{grenoble}
\and
Max Plank Institute f\"{u}r Sonnensystemforschung, Justus-von-Liebig-Weg 3, D-37077 G\"{o}ttingen, Germany
\label{maxplanck}
\and
Unidad de Astronom\'ia, Fac. Cs. B\'asicas, Universidad de Antofagasta, Avda. U. de Antofagasta 02800, Antofagasta, Chile
\label{chile}
\and
ATOS for CNES Centre Spatial de Toulouse, 18 avenue Edouard Belin, 31401 Toulouse Cedex 9, France
\label{atos}
\and
Max Planck Institute for Extraterrestrial Physics, High Energy Group, Gie{\ss}enbachstra{\ss}e, 85741 Garching, Germany 
\label{garching}
\and
Institute for Astronomy, University of Edinburgh, Royal Observatory, Blackford Hill, Edinburgh EH9 3HJ, United Kingdom
\label{edinburgh}
\and
LNE-SYRTE, Observatoire de Paris, Université PSL, CNRS, Sorbonne Université, 61 avenue de l’Observatoire 75014 Paris
\label{syrte}
\and
Faculty of Mathematics and Physics, University of Ljubljana, Jadranska ulica 19, 1000 Ljubljana, Slovenia
\label{ljubljana} 
\and
INAF-Osservatorio Astronomico di Padova, vicolo Osservatorio 5, 35122 Padova, Italy
\label{padova}
\and
INAF Astronomical Observatory of Padova, I-36012 Asiago (VI), Italy
\label{asiago}
\and
Research School of Astronomy and Astrophysics, Australian National University, Canberra, ACT 2611, Australia
\label{canberra}
}

\date{Received \textbf{Month Day, 201X}; accepted \textbf{Month Day, 201X}}

\abstract
  {For \gdrtwo{}, \spectraProcessed{} spectra, collected by the \RVS{} instrument on-board \gaia{}, were processed and median radial velocities were derived for \starProcessed{} sources brighter than \grvs~$= 12$~mag.}
  {This paper describes the validation and properties of the median radial velocities published in \gdrtwo{}.}
  {Quality tests and filters are applied to select, from the \starProcessed{} radial velocities, those with the quality to be published in \gdrtwo{}. The accuracy of the selected sample is assessed with respect to ground-based catalogues. Its precision is estimated using both ground-based catalogues and the distribution of the \gaia{} radial velocity uncertainties.}
  {\gdrtwo{} contains median radial velocities for \starPublished{} stars, with $\Teff{}$ in the range $[3550, 6900]$~K, which passed succesfully the quality tests. The published median radial velocities provide a full sky-coverage and have a completness with respect to the astrometric data of \meanCompleteness{} (for $G \leq 12.5$~mag). The median radial velocity residuals with respect to the ground-based surveys vary from one catalogue to another, but do not exceed a few 100s~$\ms$. In addition, the \gaia{} radial velocities show a positive trend as a function of magnitude, which starts around \grvs{}~$\sim 9$~mag and reaches about $+500$~$\ms$ at \grvs~$= 11.75$~mag. The overall precision, estimated from the median of the \gaia{} radial velocity uncertainties, is \medianUncertainty{}~$\kms$. The radial velocity precision is function of many parameters, in particular the magnitude and effective temperature. For bright stars, \grvs{} $\in [4, 8]$~mag, the precision is in the range 200-350~$\ms$, which is about 3 to 5~times more precise than the pre-launch specification of 1~$\kms$. At the faint end, \grvs~$= 11.75$~mag, the precisions for $\Teff =$ 5000~K and 6500~K are respectively 1.4~$\kms$ and 3.7~$\kms$.}
  {}
  

\maketitle

\section{Introduction}
ESA's \gaia{} mission \citep{GaiaPrusti2016} was launched from the Kourou space centre on 19$^{th}$ December 2013. It took about 4 weeks to the satellite to reach its operational orbit around the second Lagrange point (L2) of the Sun-Earth system. Following the commissioning phase, the nominal mission began on 25$^{th}$ July 2014, initially for a 5~years period, which has recently been extended a first time by 1.5 years\footnote{The current cold gas fuel supplies of \gaia{} would allow to extend the nominal mission by 5 years.}. \gaia{} scans continuously the celestial sphere with its 2 telescopes and its 3 instruments: the astrometric instrument, a spectro-photometer and a medium resolving power spectrograph, the \RVS{} (RVS, \citealt{DR2-DPACP-46}). The data collected are transmitted daily to Earth, when the satellite is in contact with one of the ground-based antennae. Once received on the ground, the data are processed by the \gaia{} \textit{Data Analysis and Processing Consortium} (DPAC). The consortium publishes the products of the processing in successive data releases. The first one, \gdrone{} \citep{GaiaBrown2016} was issued on 16$^{th}$ September 2016 and the second one, \gdrtwo{} \citep{DR2-DPACP-36}, on 25$^{th}$ April 2018.

Each new \gaia{} data release comes with new products. One of the novelties of \gdrtwo{} is the publication of the median radial velocities, extracted from the RVS spectra. The \RVS{} collects spectra down to \grvs~$= 16.2$~mag. Yet, because of the modest exposure time per CCD of 4.42~s, the signal-to-noise ratio of the spectra of the faintest stars is very low. The derivation of the radial velocities of the faintest stars will require to combine all the spectroscopic information collected for each source over the full mission. For \gdrtwo{}, \spectraProcessed{} spectra, recorded during the 22 first months of the nominal mission, were processed and median radial velocities were obtained for \starProcessed{} stars down to magnitude \grvs~$= 12$~mag. The spectroscopic pipeline and the processing of the spectra are described in details in a companion paper \citep{DR2-DPACP-47}.

During the last 15 years, spectroscopic surveys have delivered radial velocities\footnote{as well as stellar parameters and abundances.} for large stellar samples, e.g. Geneva-Copenhagen-Survey \citep{Nordstrom2004}: $\sim$17\,000 stars, SEGUE \citep{Yanny2009}: $\sim$240\,000 stars, APOGEE-2 \citep{Majewski2017, Abolfathi2017}: $\sim$263\,000 stars, RAVE \citep{Steinmetz2006, Kunder2017}: $\sim$460\,000 stars, Gaia-ESO-Survey \citep{Gilmore2012, Sacco2014, Jackson2015}: $\sim$50\,000 stars in DR3\footnote{\url{https://www.gaia-eso.eu/news/archive/public-data-release-3}}, LAMOST \citep{Cui2012, Zhao2012}: $\sim$5.3~million stars in DR5\footnote{\url{dr5.lamost.org}}, HERMES-GALAH \citep{Martell2017}: $\sim$200~000 stars. In the continuity of these large surveys, \gdrtwo{} contains median radial velocities for \starPublished{} stars with $\Teff{}$ in the range $[3550, 6900]$~K and distributed over the full celestial sphere.

This paper is devoted to the description and validation of the median radial velocities published in \gdrtwo{}. After a short summary of the main characteristics of the \RVS{} instrument (Sect.~\ref{sec:RVS}) and a brief overview of the spectroscopic processing pipeline (Sect.~\ref{sec:pipeline}), Sect.~\ref{sec:filters} presents the filters that were applied after the completion of the processing, to select the radial velocities with the quality to be published in \gdrtwo{}. Sect.~\ref{sec:dr2cat} describes the properties of the published radial velocities.


\section{The Radial Velocity Spectrometer\label{sec:RVS}}
The \emph{Radial Velocity Spectrometer} (RVS) is described in details in \citep{DR2-DPACP-46}. Here we provide a very brief description of the main characteristics the instrument.

The {\it Radial Velocity Spectrometer} is a medium resolving power, $R = \lambda / \Delta \lambda = 11\ 500$, near-infrared $\lambda \in [845, 872]$~nm, integral field spectrograph. As the astrometric and photometric instruments, the RVS is illuminated by the two Gaia telescopes. The spectra are recorded on a block of 12 CCDs\footnote{3 in the direction of the scan times 4 in the direction perpendicular to the scan} located at the edge of the Gaia focal plane. When a source crosses one of the two fields of view (hereafter referred to as {\it transit}), 3 spectra are recorded, i.e. one per CCD along the scan direction. The exposure time per CCD is 4.42~s. On average, the RVS should record 40 transits per source during the first 5 years of the mission.

Gaia is continuously spinning and scanning the sky with a 6 hours period. The CCDs are therefore operated in {\it Time Delay Integration} (TDI) mode, i.e. the charges are transferred from columns to columns at high frequency, in order to follow the sources during their crossing of the focal plane. In order to minimize both the telemetry budget and the electronic noise, elongated windows are read around the spectra and transmitted to the ground. The rest of the pixels are flushed in the readout register and discarded. At the start of the nominal mission, the windows were 1260 pixels long\footnote{i.e. in the sense of the spectral dispersion which is also the orientation of the scan} and 10 pixels wide. The length of the windows was increased to 1296 pixels in Spring 2015 to allow for a better measure of the background light. The on-board software allocates windows to sources down to \grvs = 16.2~mag, which is the limiting magnitude of the RVS.

The RVS is an integral field spectrograph and, as such, disperses the light of all the sources contained in its two fields of view, with the consequence
that the spectra of very close neighbours will overlap. In this case, to avoid transmitting twice the pixels, the on-board software can decide to truncate the windows in the direction perpendicular to the dispersion, i.e. allocating less than 10 pixels to the window width. If the conflict does not extend on the full length of the spectrum, the truncation will be applied only to the appropriate portion of the spectrum, resulting in a non-rectangular window. For Gaia DR2, only rectangular windows were processed. The specific treatments required by non rectangular windows are planned for Gaia DR3.

\begin{figure}[h!]
\centering
\includegraphics[width=0.54\textwidth]{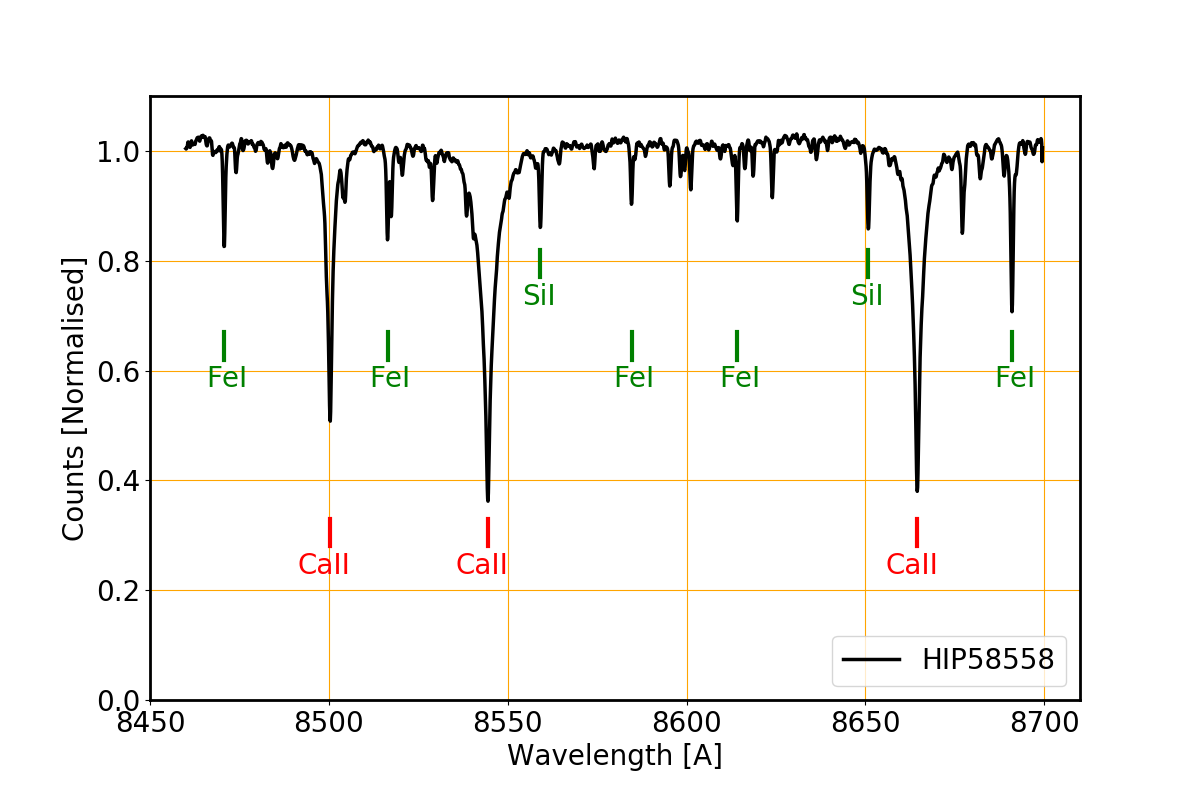}      
\caption{RVS spectrum of the star HIP~58558.}
\label{fig:hip58558}
\end{figure}

In \gdrtwo{}, median radial velocities are published for stars with $\Teff$ in the interval $[3550, 6900]$~K. In this temperature range, the RVS spectra are dominated by a triplet of the ionised calcium. The wavelength range also contains weaker neutral metallic lines of, e.g. iron, silicon or titanium. As an example, Fig.~\ref{fig:hip58558} presents the RVS spectrum of the star HIP~58558: $\Teff = 5477$~K, $\logg = 4.34$ and $\FeH = 0.02$~dex \citep{Adibekyan2012}. Below about 3400~K, the RVS spectra develop strong molecular bands. In A and B spectral types, the P13 to P17 hydrogen Paschen lines become the dominant spectral features. As discussed in Sect.~\ref{sec:filters}, the radial velocities of \emph{cool} and \emph{hot} stars, i.e. outside $[3550, 6900]$~K, will be published in a future \gaia{} release.

\section{The spectroscopic pipeline\label{sec:pipeline}}
\subsection{Overview}
The \gaia{} spectroscopic processing pipeline is described in details in \citet{DR2-DPACP-47}. In this section, we provide a brief overview of its functionalities.\\

The spectroscopic pipeline is in charge of four main tasks:
\begin{enumerate}
\item {\it Calibrating the RVS instrument.} The pipeline developed for \gdrtwo{} calibrates: the electronic bias, the read-out noise, the wavelengths and the \grvs magnitude zero-point. For the other RVS characteristics, the pipeline relies either on pre-launch or off-line calibrations.
\item {\it Reducing and cleaning the spectra.} This includes in particular, e.g. subtracting the electronic bias and the straylight, discarding spectra degraded by CCD cosmetic defects or saturation, applying the wavelength calibration, removing the cosmic rays, normalising the continuum, deriving the internal \grvs{} magnitude, selecting a template to derive the radial velocity. 
\item {\it Measuring the radial velocities per transit.} For each star and for each transit one (or two, in the case of a double star) radial velocity is measured. Each radial velocity is derived by a series of modules which compare the 3 RVS spectra collected per transit to one (or several) template(s) shifted step by step in radial velocity, from $-$1000 to $+$1000~\kms. This workflow is hereafter referred to as \emph{Single Transit Analysis (STA)}.
\item {\it Combing the information of the successive transits.} For each source, the radial velocities derived on successive transits are grouped in a time series. The radial velocities published in \gdrtwo{} are the medians of these time series (see Sect.~\ref{sec:vr} for more details). Statistical properties of the time series are also calculated and are used as variability and quality indicators. In addition, for each source, all the spectra recorded are shifted to rest frame and combined. The combined spectra are then examined in order to detect possible emission lines. This set of tasks is hereafter referred to as \emph{Multiple Transit Analysis (MTA)}.
\end{enumerate}

\subsection{The \grvs{} magnitudes\label{sec:grvs}}
Different measures of the \grvs{} magnitude are either produced or used by the pipeline and in this paper. They are defined below:
\begin{itemize}
\item The {\it On-board \grvs} (\obgrvs{}) is derived by the Gaia on-board software.
\item The {\it External \grvs} (\extgrvs{}) is calculated from ground-based photometric catalogues using colour-colour transformations. When no ground-based photometry is available, the on-board \obgrvs{} is adopted. \extgrvs{} is the magnitude used to define the limiting magnitude of \gdrtwo{}: \extgrvs$\leq 12$~mag. 
\item The {\it Internal \grvs} (\intgrvs{}) is measured by the spectroscopic pipeline from the RVS spectra. For the faint stars, the accuracy and precision of \intgrvs{} are limited by the basic modelling of the straylight in \gdrtwo{}. A more elaborated calibration of the straylight is in development for \gdrthree{}.
\end{itemize}


\section{Selecting radial velocities for publication in \gdrtwo{}  \label{sec:filters}}
For \gdrtwo{}, the CU6 pipeline has processed \spectraProcessed{} spectra and produced radial velocities for \starProcessed{} stars without pre-selection on spectral type or colour indices and for a very broad range of signal to noise ratios. The pipeline includes validation functionalities, described in \citet{DR2-DPACP-47}, which can identify autonomously and reject problematic data: e.g. negative total spectrum flux or nearby duplicated transits. These diagnostics are usually meant to detect issues at spectrum or transit level. They do not consider the global properties of the data (known a-posteriori) and the potential outliers. Therefore, following the completion of the processing, an off-line validation campaign was conducted on the full \starProcessed{} stars sample to assess its characteristics and identify the stars which did not had the quality to be published in \gdrtwo{}. This resulted in the following list of filters:\\

{\it Large coordinates uncertainties.} The right ascensions and declinations (computed by the \gaia{} astrometric pipeline, see \citealt{DR2-DPACP-51}) are used to derive the coordinates of the sources in the RVS field of views, which are then used to calibrate the spectra in wavelength. The uncertainties on the coordinates of the sources are therefore propagated to the wavelengths of the spectra and {\it in fine} to the radial velocities. In \gdrtwo{}, the mean precision on the sources positions is 0.03~mas, which represents a very minor contribution to the radial velocity error budget of $\sim 4.3$~m/s. Of course, a small fraction of the stars presents much larger astrometric uncertainties. The radial velocities of stars with a quadratic sum of the uncertainties on the right ascension and on the declination, i.e. $\sqrt{\epsilon_\alpha^2 + \epsilon_\delta^2}$, larger than 100~mas (corresponding to $\sim 14.5$~km/s) were discarded from \gdrtwo{}.\\

{\it Faint stars.} For \gdrtwo{}, stars brighter than \extgrvs{} = 12~mag were processed by the spectroscopic pipeline. The selection was performed using the external Grvs magnitude (see Sect.~\ref{sec:grvs}). The spectroscopic pipeline also derives an internal \intgrvs{} magnitude based on the flux contained in the RVS spectrum between 846 and 870~nm. It was considered that stars with an internal \intgrvs magnitude equal to or fainter than 14~mag were not containing enough signal per spectrum to yield a good enough velocity in \gdrtwo{}.\\

{\it Ambiguous transits.} The \emph{Single Transit Analysis (STA)} workflow produces a boolean quality flag, {\it isAmbiguous}, which identifies the radial velocities which look suspicious on a specific transit. In validation, the ratio of the number of {\it ambiguous} transits over the total number transits, for each source, was examined. The distribution on the celestial sphere of this new quality indicator shows a very specific pattern. While on most of the sky, the mode of its distribution peaks at 0, in a few specific areas it peaks at 1 (i.e. peaks at 100\% of ambiguous transits per source). This behaviour seems, for a part, related to the overestimation of the external \extgrvs{} magnitude in some specific areas, allowing some faint stars to enter the spectroscopic processing. The radial velocities of stars with 100\% ambiguous transits were excluded from \gdrtwo{}. The two filters on {\it faint stars} and {\it ambiguous transits} show some overlap, both rejecting stars with too low signal in the RVS window. Yet, because they consider different quantities, they are also complementary, one possibly identifying a suspicious star that the other could have missed.\\

{\it Large radial velocity uncertainties.} The distribution of unfiltered radial velocity uncertainties peaks between 1 and 2~km/s and shows either an extended tail or a broad small amplitude secondary peak (depending on the location on the sky). The extended tail/secondary peak includes a mix of stars with insufficient signal to be processed in \gdrtwo{}, large amplitude variables and undetected binary or multiple systems. In all these cases, the median radial velocity published in \gdrtwo{} would not be a reliable estimate of the source or system radial velocity. The median radial velocities of stars with a radial velocity uncertainty larger or equal to 20~km/s were discarded from \gdrtwo{}.\\

{\it Suspected double lines spectroscopic binaries.} The \emph{Single Transit Analysis} workflow includes a software module which is in charge of detecting the spectra presenting double line patterns and to derive their two radial velocities. The multi-instrument modelling and publication of binary systems is planned for \gdrthree{}. Therefore, in \gdrtwo{}, stars with more than 10\% of their transits flagged as double lines patterns were considered as potential double-lines spectroscopic binaries (SB2) and their median radial velocities were removed from \gdrtwo{}.\\

\begin{figure}[h!]
\centering
\includegraphics[width=0.54\textwidth]{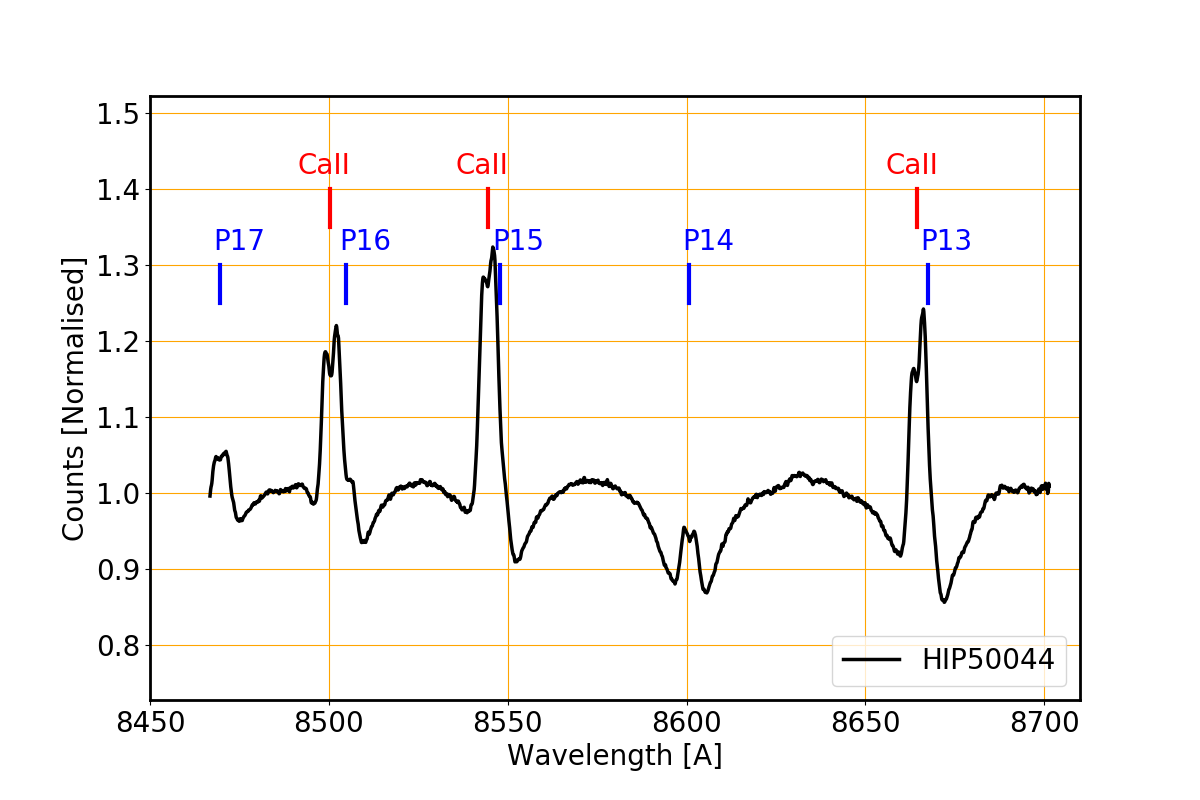}
\caption{RVS spectrum of the Be star HIP~55044. Both the \ion{Ca}{ii} and the Paschen lines are in emission.}
\label{fig:hip55044}
\end{figure}

{\it Suspected emission lines stars.} The \emph{Multiple Transit Analysis (MTA)} workflow includes a software module in charge of detecting emission line stars. The library of spectra used for \gdrtwo{} does not contain emission lines templates. The comparison of an emission lines star with an inappropriate absorption lines template can produce systematic radial velocity shifts of several hundreds \kms{}. The radial velocities of stars identified as potential emission lines were excluded from \gdrtwo{}. Figure~\ref{fig:hip55044} shows, as an example, the RVS spectrum of the star HIP~55044, which has been detected as an emission lines star by the \emph{MTA} workflow. It is classified in the literature as a Be star \citep{Houk1975}.\\

{\it Cool stars.} In the RVS domain, the spectra of late M-stars are dominated by TiO molecular bands. The spectra show steep slopes and because of the narrow RVS wavelength range, the pseudo-continuum is usually not visible on the side of the shortest wavelengths. For these stars, the radial velocity precision is more sensitive, than other for spectral type, to the treatment of the continuum as well as to the choice of the template. Tests conducted during the validation phase showed that the radial velocity performance were significantly lower for the stars processed using templates with effective temperatures of 3500~K or lower. The radial velocities of these stars were removed from \gdrtwo{}. The spectroscopic pipeline will be upgraded to process late M-type stars and publish their radial velocities in a future \gaia{} data release.\\

{\it Hot stars.} About 18\% of the stars processed for \gdrtwo{} had known ground-based atmospheric parameters. When available, this information was used to select the closest template in a library of several thousands synthetic spectra. When the parameters were unknown, a dedicated software module, \emph{determineAP}, was in charge of choosing the template by comparison of the RVS spectrum to a set a synthetic spectra \citep{DR2-DPACP-47}. For \gdrtwo{}, \spectraProcessed{} spectra were processed. The volume of data was too large to compare 85\% of the RVS spectra to the full library of synthetic spectra. To keep the processing load reasonable, \emph{determineAP} had to select the templates in a sub-grid restricted to 28 synthetic spectra, with in particular a single choice of surface gravity per effective temperature. The validation phase showed that this was insufficient for hot stars. In particular, in early F and A-type stars, the Paschen lines are significantly pressure sensitive and get stronger with decreasing surface gravity. Three of the Paschen lines, i.e. P13, P15 and P16, are blended with the \ion{Ca}{ii} lines. The profiles of the blended lines change with surface gravity and as a consequence, the centroids of the lines are shifted. This can produce a systematic bias on the radial velocity of several \kms{}, if it is derived with the inappropriate template. The systematic error can be amplified if the star shows a significant projected rotational velocity, which will modify more the "narrow" profile of the \ion{Ca}{ii} lines than the broader profiles of the Paschen lines. Therefore, the radial velocities derived using a template with an effective temperature of 7~000~K or higher were excluded from \gdrtwo{}.

\begin{figure}[h!]
\centering
\includegraphics[width=0.54\textwidth]{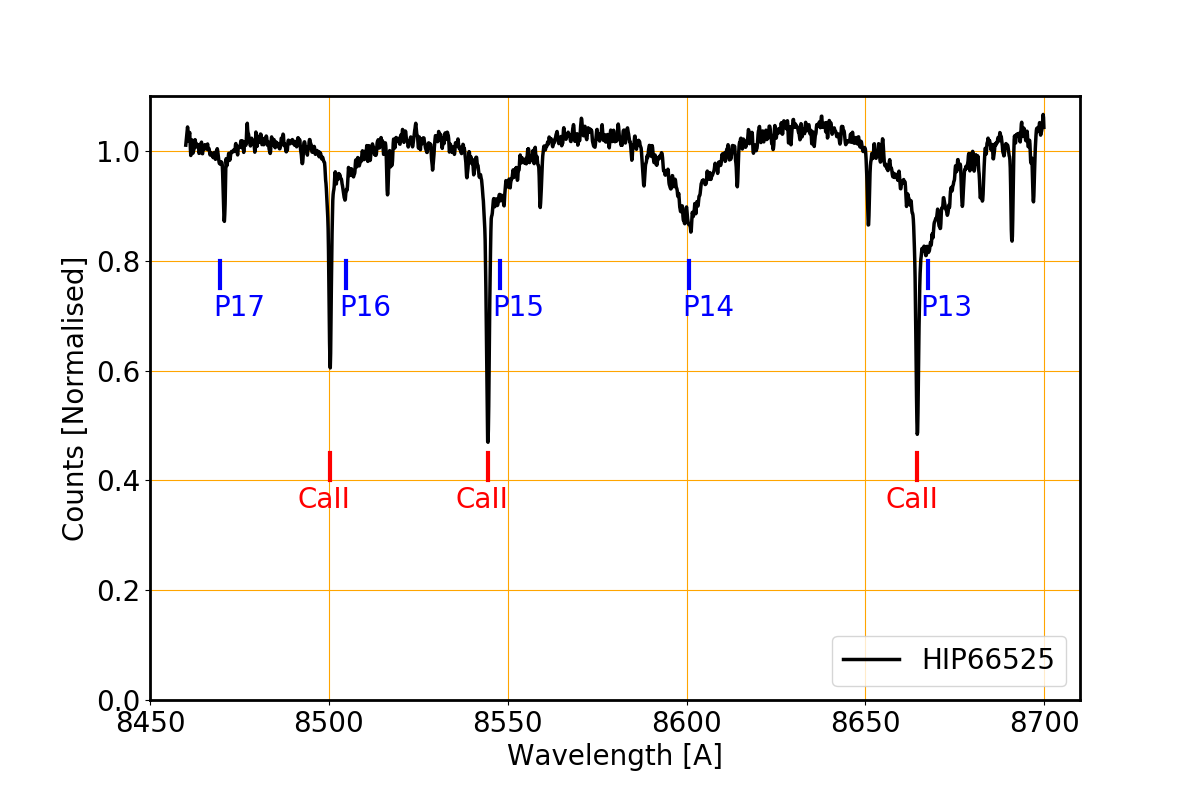}
\includegraphics[width=0.54\textwidth]{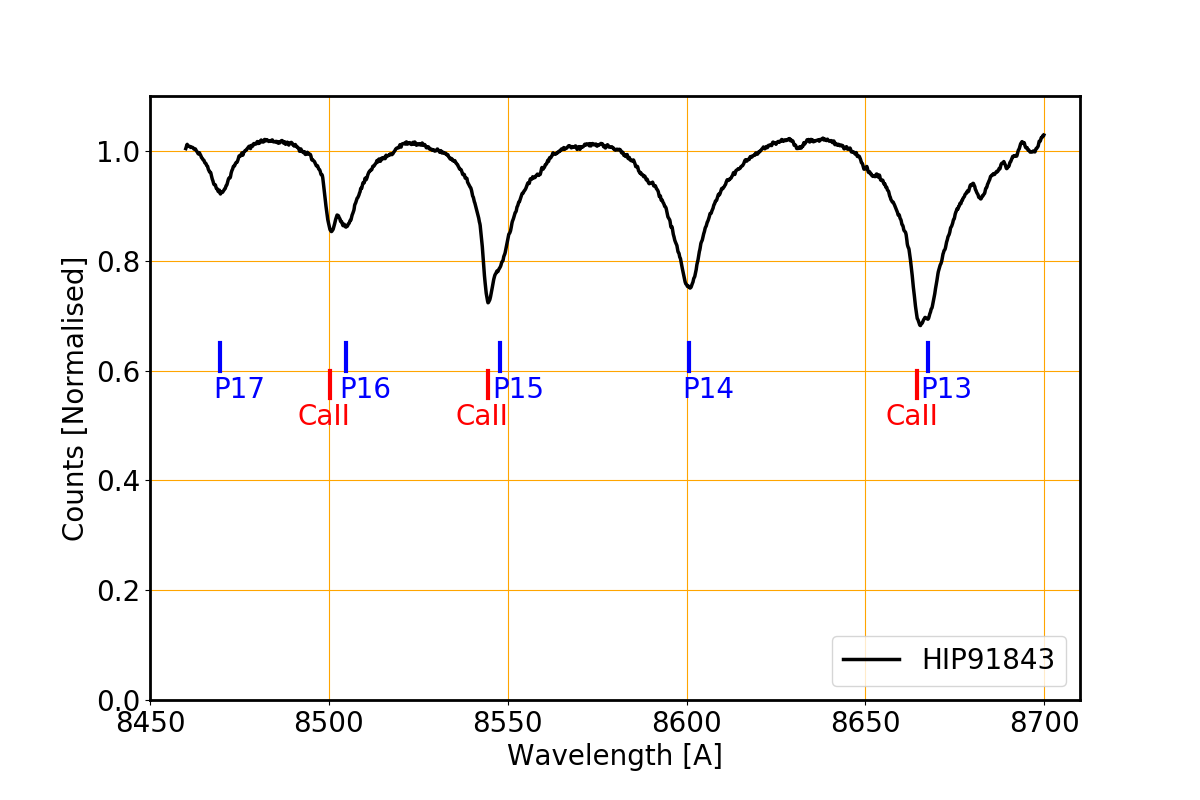}      
\caption{RVS spectra of HIP~66525 (top) and HIP~91843 (bottom).}
\label{fig:hip66525}
\end{figure}

Figure~\ref{fig:hip66525} illustrates the sensitivity of the \ion{Ca}{ii}-Paschen blends to surface gravity and rotational velocity. HIP~66525 (top) has a $\Teff \sim 8000$~K and a $\logg \sim 4.6$ \citep{Kordopatis2013}, while HIP~91843 (bottom) has a $\Teff \sim 8100$~K and a $\logg \sim 3.2$ (derived from 2MASS photometry). One can see that the Paschen lines are stronger in the latter than in the former. Moreover, in HIP~91843, the depth of the \ion{Ca}{ii} lines is reduced by the projected rotational velocity broadening.

In \gdrthree{} and following releases, the atmospheric parameters should be derived by elaborated analysis of the \gaia{} spectro-photometric data and, for the brightest stars, of the RVS spectra. The spectroscopic pipeline will also be upgraded to derive the rotational velocities. The radial velocities of hot stars will be published when the precisions on the atmospheric parameters and on the rotational velocities, will allow to process them with the appropriate template and blends profiles.\\

{\it High-velocity stars.} Of the more than 7 million stars that passed successfully the above filters, 613 had absolute radial velocities larger or equal to 500~\kms. Because, this sample of high-velocity stars is very small compared to the full sample it can easily be significantly contaminated by outliers. For example, an \emph{hypothetical} rate of 1 undetected outlier out of 10~000 stars, generating a random (and wrong) radial velocity, would produce about 350 false high-velocity stars. As a consequence, a particular attention was paid to them. Their combined spectra (see Sect.~\ref{sec:pipeline}) were visually examined one by one and the proper locations of their \ion{Ca}{ii} lines was checked. Out of the 613 stars, 216 were considered as valid high-velocity stars and the median radial velocity of the 397 others were removed from \gdrtwo{}. Another 14 stars were rejected by astrometric and/or photometric filters, so that \gdrtwo{} contains 202 stars with $|V_R| \geq 500$~$\kms$. A-posteriori, it is possible to assess the contamination rate before cleaning. It was decreasing with the absolute value of the radial velocity and was of $\sim90$\% in the range $[950, 1000]$~\kms and $\sim20$\% in the range $[500,550]$~\kms. The number of stars was raising too fast with decreasing absolute value of the velocity to extend the systematic visual inspection to slower stars. It is therefore important, when working with \gdrtwo{} high-velocity stars, to take into account that an additional quality filter has been applied to stars faster than 500~\kms, resulting in different selection functions for the stars below and above this value.\\

In total, 2.6 million median radial velocities were discarded for \gdrtwo{}, mostly by the above spectroscopic filters, but some also by other DPAC filters based on photometric or astrometric criteria \citep{DR2-DPACP-36, DR2-DPACP-51, DR2-DPACP-40, DR2-DPACP-39}.\\

\section{The \gdrtwo{} radial velocity catalogue\label{sec:dr2cat}}
\subsection{Catalogue content\label{sec:content}}
\gdrtwo{} contains median radial velocities for \starPublished{}~stars as well as their radial velocity uncertainties, number of transits and template parameters (effective temperature, surface gravity and metallicity). The spectroscopic fields published in \gdrtwo{} are listed in Table~\ref{tab:dr2Fields}.

\begin{table}[h!]
\caption[]{Spectroscopic content of Gaia-DR2.\label{tab:dr2Fields}}
\begin{tabular}{| l | l | l |} \hline
Field                       & Units    & DB column name          \\ \hline
Median radial velocity      & \kms     & radial\_velocity        \\
Radial velocity uncertainty & \kms     & radial\_velocity\_error \\
Number of transits          & transits & rv\_nb\_transits        \\
Template temperature        & K        & rv\_template\_teff      \\
Template surface gravity    & dex      & rv\_template\_logg      \\
Template metallicity        & dex      & rv\_template\_fe\_h     \\ \hline
\end{tabular}
\end{table}

\begin{figure}[h!]
\centering
\includegraphics[width=0.5\textwidth]{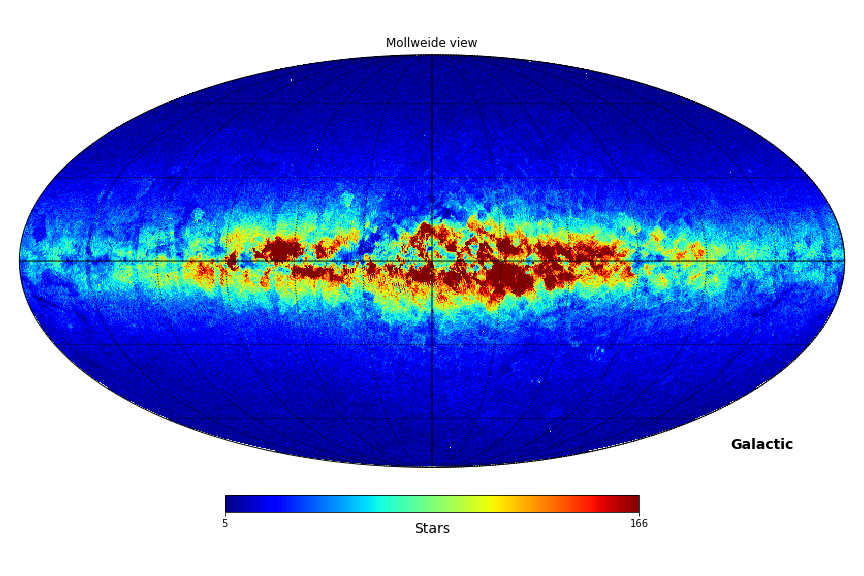}      
\caption{Distribution on the sky of the \starPublished{}~stars with a radial velocity in \gdrtwo{}. The projection is in Galactic coordinates. The Galactic Centre is in the middle of the figure and the galactic longitudes increase to the left. The pixel size is about 0.2 square degree (healpix level 7).}
\label{fig:counts}
\end{figure}

The spectroscopic catalogue has a full sky coverage. Figure~\ref{fig:counts} shows the distribution in Galactic coordinates of the stars with a radial velocity in \gdrtwo{}. The vast majority of the stars belong to the Milky-Way, but some bright Magellanic Clouds members are also part of this release.

The external \extgrvs, used in \gdrtwo{} to define the limiting magnitude of the spectroscopic pipeline, is mainly calculated from ground-based photometry (Sect.~\ref{sec:grvs}). Different catalogues were used in different areas of the sky and with different colour-colour relations. This led to small variations of the \extgrvs zero-point and therefore of the limiting magnitude of the processing, as a function of celestial coordinates. The variations are visible in the star counts as small amplitudes (few) degree(s) scale patterns.

\begin{figure}[h!]
\centering
\includegraphics[width=0.5\textwidth]{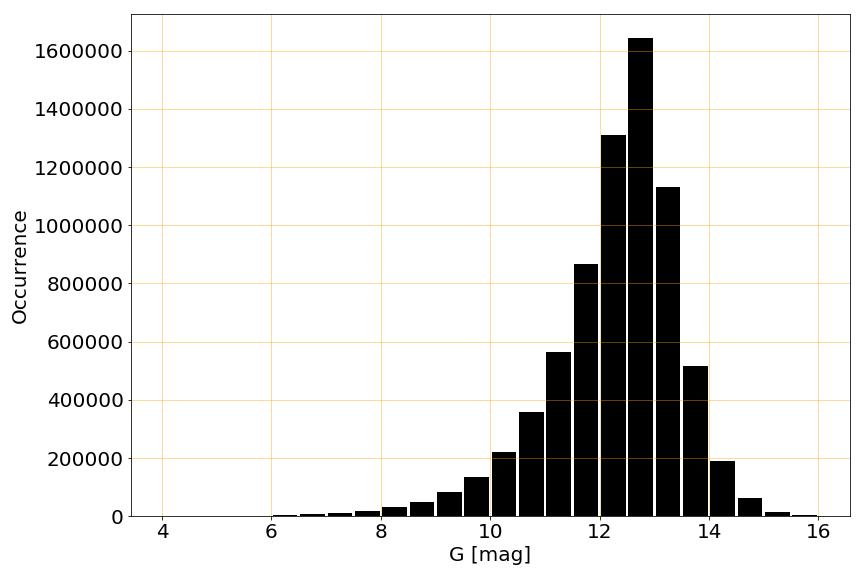}      
\caption{$G$-magnitude distribution of the stars with a radial velocity published in \gdrtwo{}.}
\label{fig:histoG}
\end{figure}

Figure~\ref{fig:histoG} shows the $G$-magnitude distribution of the stars with a radial velocity in \gdrtwo{}. 

\begin{figure}[h!]
\centering
\includegraphics[width=0.5\textwidth]{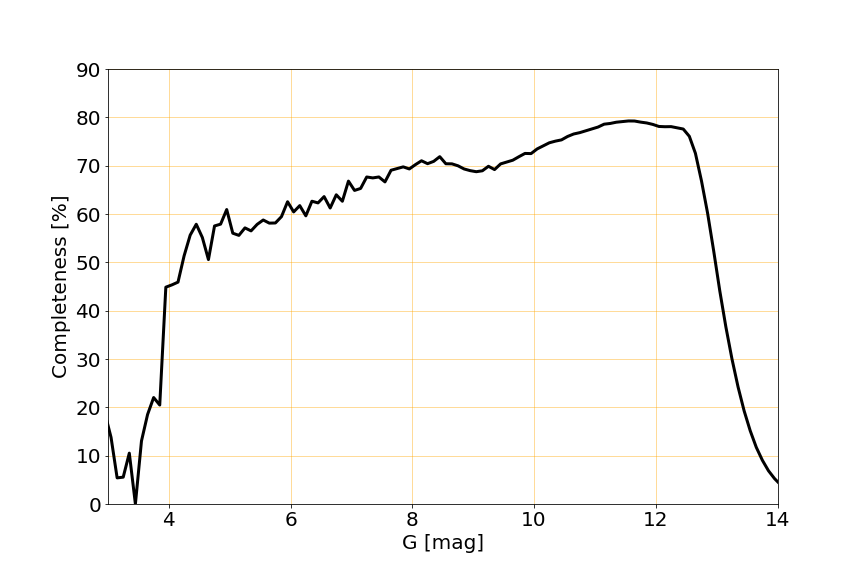}      
\caption{Completeness with respect to the full \gdrtwo{} catalogue, as a function of G magnitude.}
\label{fig:compl}
\end{figure}

Figure~\ref{fig:compl} shows the completeness of the \gdrtwo{} radial velocities with respect to the full \gdrtwo{} catalogue, as a function of $G$-magnitude. The completeness increases smoothly from $G \sim 4$ to $G \sim 11.5$~mag. The steep decrease for the faint stars is the consequence of the limiting magnitude of \gdrtwo{} processing: \extgrvs{}$= 12$~mag. The sharp cut-off at $G \sim 4$~mag is due to the saturation of the core of the RVS spectra, which are then discarded by the spectroscopic pipeline. The completeness for $G \leq 12.5$ is 77.2\%.

\begin{figure}[h!]
\centering
\includegraphics[width=0.5\textwidth]{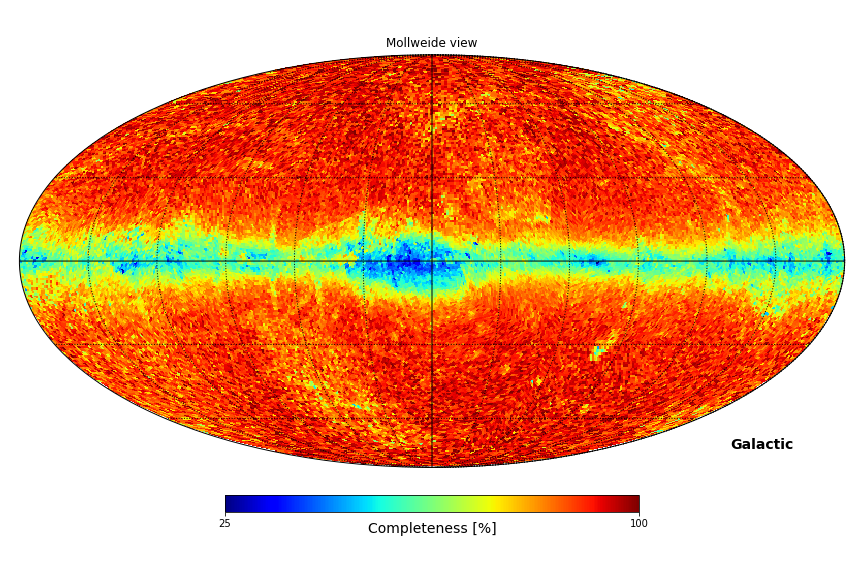}      
\caption{Completeness with respect to the \gdrtwo{} catalogue, for the stars with $G \leq 12.5$~mag, as a function of galactic coordinates. The Galactic Centre is in the middle of the figure and the galactic longitudes increase to the left. The pixel size is $\sim$0.8 square degree (healpix level 6).}
\label{fig:complMap}
\end{figure}

Figure~\ref{fig:complMap} shows the completeness with respect to the \gdrtwo{} catalogue, as a function of galactic coordinates, for the stars with $G \leq 12.5$~mag. At the first order, the completeness is driven by the projected stellar density. In dense areas, the conflicts between RVS windows are more frequent, leading to a higher probability of the windows to be truncated and therefore not processed by the pipeline. As a consequence, the completeness is lower in the directions of the Galactic bulge and in the Galactic disc and rapidly increases as one moves away from the Galactic plane.

\subsubsection{Median radial velocity\label{sec:vr}}
The radial velocity published in \gdrtwo{} is the median of the radial velocities derived per transit. Some observations were not used for the calculation of the median:
\begin{itemize}
\item Truncated windows (caused by the overlap with the window of another star) were not processed for \gdrtwo{} and therefore had no transit Vr derived (for that transit).
\item Transits for which a spectrum was flagged as double-lines spectroscopic binaries (SB2) were excluded from the calculation of the median.
\end{itemize}

A minimum of two eligible (i.e. non-truncated, non-SB2) transits was required to derive the median radial velocity of a star.\\

\begin{figure}[h!]
\centering
\includegraphics[width=0.5\textwidth]{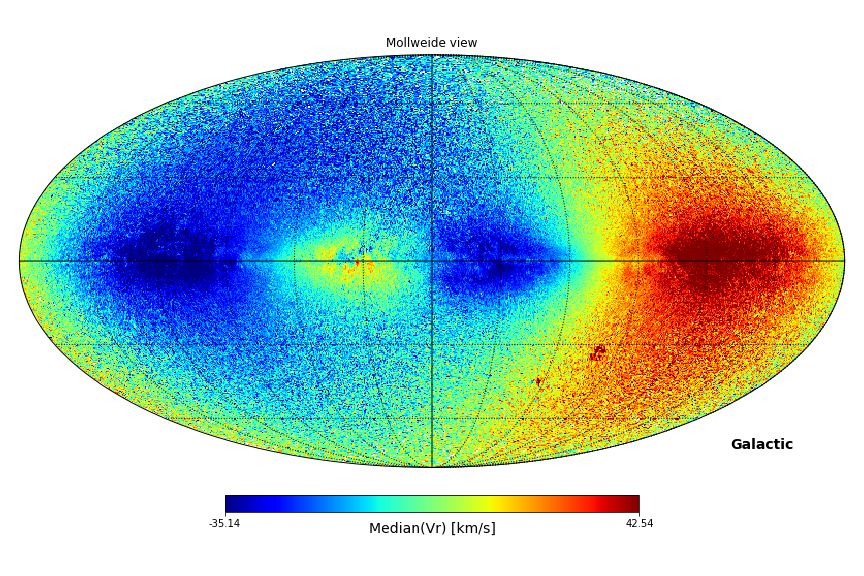}      
\caption{Median radial velocity map as a function of galactic coordinates. The Galactic Centre is in the middle of the figure and the galactic longitudes increase to the left. The pixel size is $\sim 0.2$ square degree (healpix level 7).}
\label{fig:mapsMedVr}
\end{figure}

\begin{figure}[h!]
\centering
\includegraphics[width=0.5\textwidth]{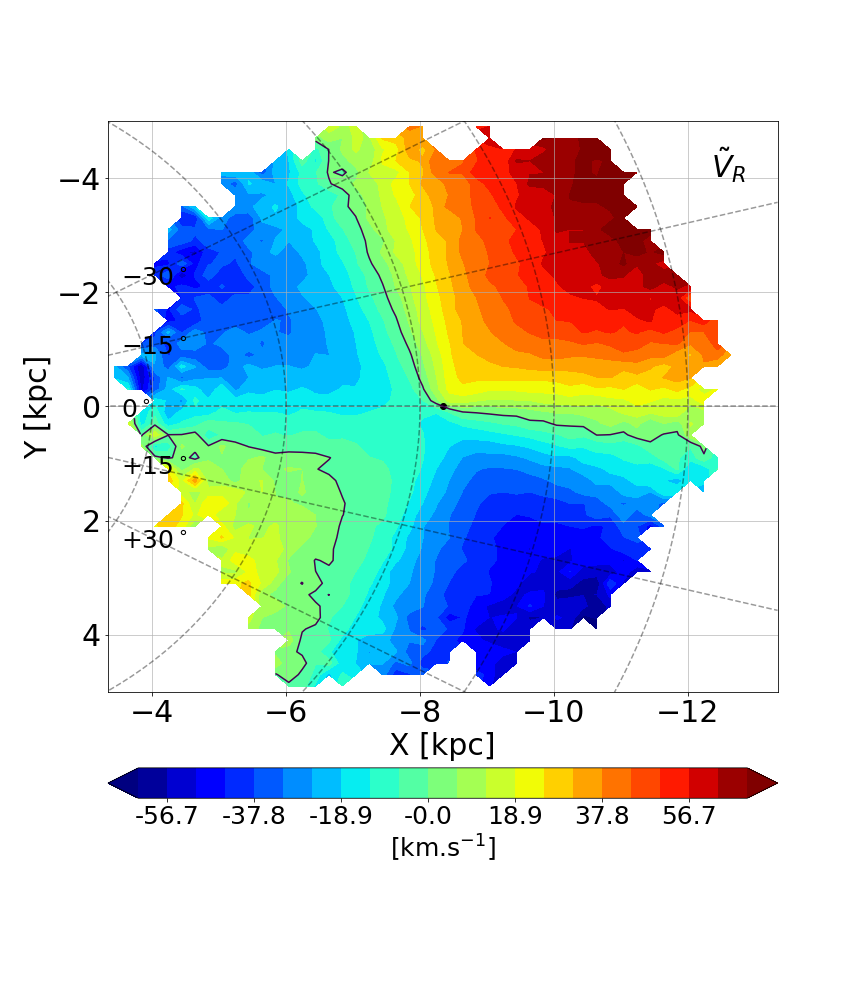}      
\caption{Face-on view map of the median radial velocity of the Galactic disc, as seen from the north Galactic pole. The Galactic azimuths are increasing clockwise. They are labelled from $-30$ to $+30$~degrees, on the left of the map. The Sun is represented by a black dot, located \emph{"arbitrarily"} at $X=-8.34$~kpc and $Y=0$~kpc. The galactic centre is located on the left side. The Milky Way is rotating clockwise. The iso-velocity contour $\tilde{V}_R = 0$ is pointed out as black lines. The map has been calculated using 5\,020\,596 stars, selected in a 2~kpc horizontal layer centred on the Galactic mid-plane.}
\label{fig:faceonMedVr}
\end{figure}

Figure~\ref{fig:mapsMedVr} shows the map of the median of the radial velocities as a function of galactic coordinates. The median of the radial velocities is calculated over healpix level 7 pixels of about 0.2~square degree each (the same area as the star counts map: Fig.~\ref{fig:counts}). Figure~\ref{fig:faceonMedVr} presents the face-on view map of the median of the radial velocities of the Galactic disc stars, for sources located within $\pm$1~kpc of the Galactic mid-plane\footnote{the stars in the face-on map were also selected on the basis of their relative parallax uncertainty: $\sigma_\varpi / \varpi \leq 20$\%.}. The median radial velocity is calculated over cells of 200~pc by 200~pc. Both maps show the line-of-sight-projected differential rotation of the stars of the Galaxy, as observed from the Sun. The Large and Small Magellanic Clouds also stands out clearly in Fig.~\ref{fig:mapsMedVr} around $(l,b) \sim (-80^\circ, -33^\circ)$ and $(-57^\circ,-44^\circ)$ respectively.

\subsubsection{Radial velocity uncertainty \label{sec:vrError}}
The radial velocity uncertainty is calculated as the uncertainty on the median of the transit radial velocities quadratically summed with a constant term of 0.11 $\kms$ which represents the current calibration noise floor:
\begin{equation}
\epsilon_{V_R} = [(\sqrt{{\pi}\over{2 N}} \sigma_{V_R^t})^2 + 0.11^2]^{0.5}
\end{equation}
where $N$ is the number of eligible transits used to derive the median radial velocity and $\sigma_{V_R^t}$ the standard deviation of the eligible transit radial velocities.

\begin{figure}[h!]
\centering
\includegraphics[width=0.5\textwidth]{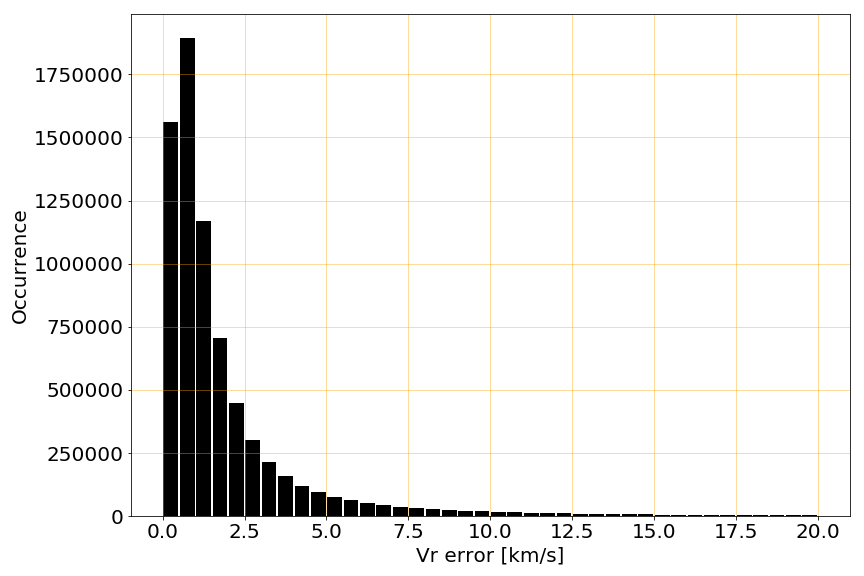}      
\caption{Distribution of the radial velocity uncertainties.}
\label{fig:vrErrorHisto}
\end{figure}

Figure~\ref{fig:vrErrorHisto} shows the distribution of the radial velocity uncertainties. The first quartile, median and third quartile of the distribution are respectively: 0.55, 1.05 and 2.08~$\kms$.

The radial velocity uncertainty relies on the standard deviation of the time series of transit radial velocities. As described in the next section (Sect.~\ref{sec:transit}), some time series are made of a few transits, in which cases the standard deviations and therefore the radial velocity uncertainties are less precise than for larger number of transits.

\subsubsection{Number of transits\label{sec:transit}}
\begin{figure}[h!]
\centering
\includegraphics[width=0.5\textwidth]{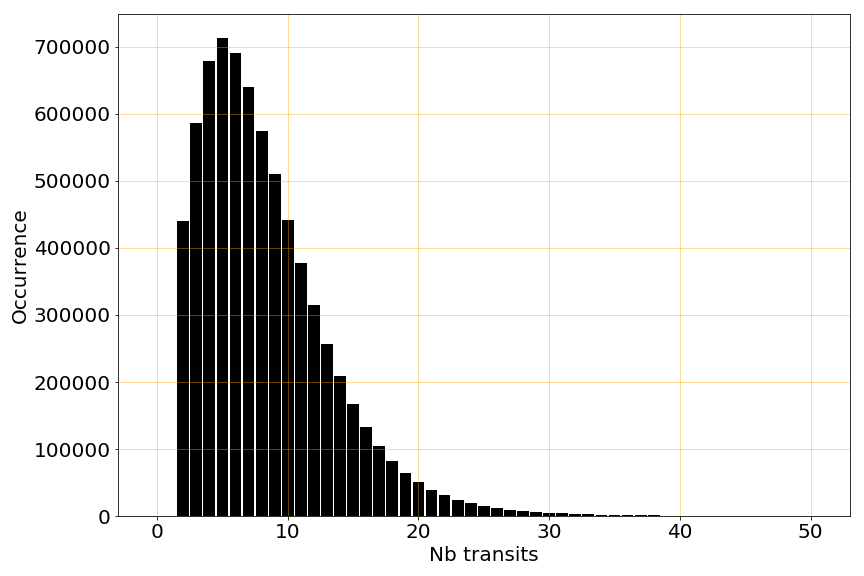}      
\caption{Distribution of the number of transits per star used to derive the median radial velocity.}
\label{fig:transitHisto}
\end{figure}

The number of transits published in \gdrtwo{} , is the number of eligible transits used to compute the median radial velocity. Figure~\ref{fig:transitHisto} shows the distribution of the number of transits. It ranges from 2 (by construction) to 201, with a median number of 7.

\begin{figure}[h!]
\centering
\includegraphics[width=0.5\textwidth]{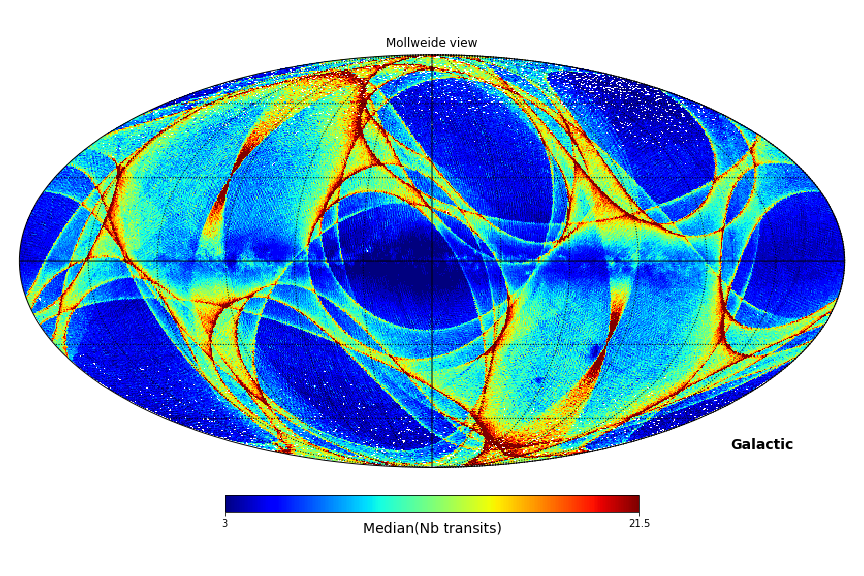}      
\caption{Distribution of the median number of transits as a function of galactic coordinates. The Galactic Centre is in the middle of the figure and the galactic longitudes increase to the left. The pixel size is about 0.2 square degree (healpix level 7).}
\label{fig:transit}
\end{figure}

The main driver of the number of observations for a source is the satellite scan law, which defines how many times a specific area of the sky has been monitored. During the first 28 days of the nominal mission, \gaia{} was in {\it "Ecliptic Pole Scanning Law"} (EPSL) and was observing each Ecliptic Pole with each telescope every 6 hours (the spin period of the satellite). The stars with large number of transits are stars close to the Ecliptic Poles which have been repeatedly monitored during the 28 days of EPSL. The second factor which defines the number of transits is the stellar density. In dense areas, the stars are closer and the probability of conflict (i.e. overlap) between RVS windows is higher. For \gdrtwo{}, overlapped windows have not been processed and as a consequence the mean number of transits is lower in dense areas. It should be noted that the satellite scans have different orientations at different transits. Because of the very elongated geometry of the windows, it is not systematic that a star overlap with the same neighbour at each transit. Therefore, the stellar density impacts first the number of transits before the completeness (which is also impacted, as shown in Sect.~\ref{sec:content}). Figure~\ref{fig:transit} shows the distribution on the sky (galactic coordinates) of the median number of transits per $\sim$0.2 square degree pixel.

\subsubsection{Parameters of the template}
\begin{figure}[h!]
\centering
\includegraphics[width=0.5\textwidth]{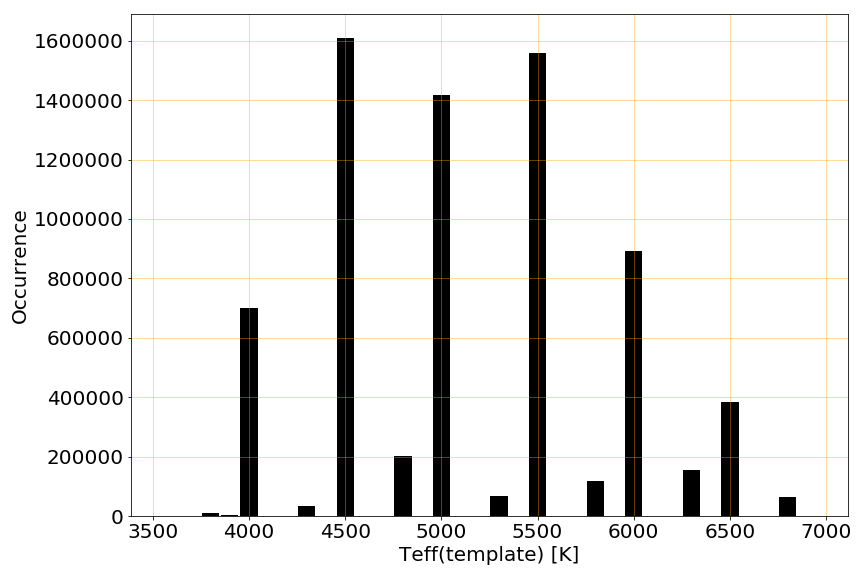}      
\caption{Distribution of the effective temperatures of the templates.}
\label{fig:tempTHisto}
\end{figure}

\begin{figure}[h!]
\centering
\includegraphics[width=0.5\textwidth]{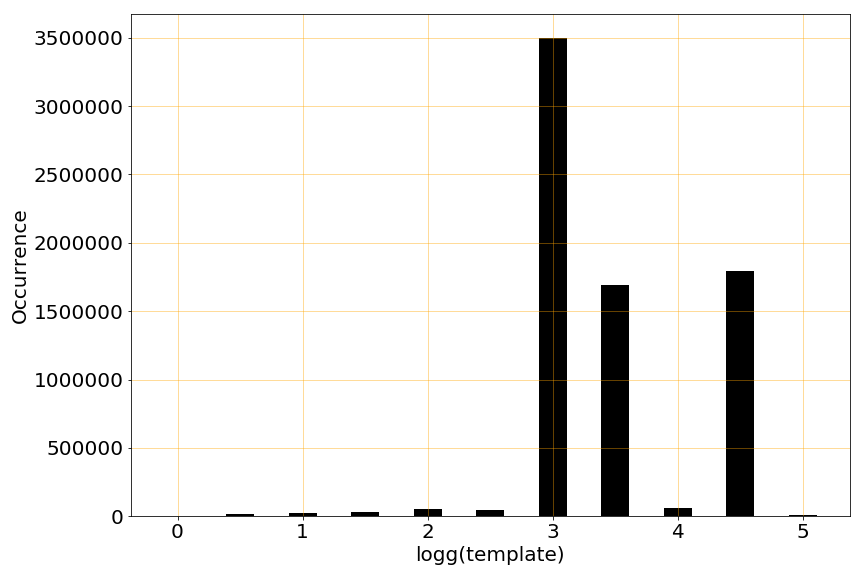}      
\caption{Distribution of the surface gravities of the templates.}
\label{fig:tempGHisto}
\end{figure}

\begin{figure}[h!]
\centering
\includegraphics[width=0.5\textwidth]{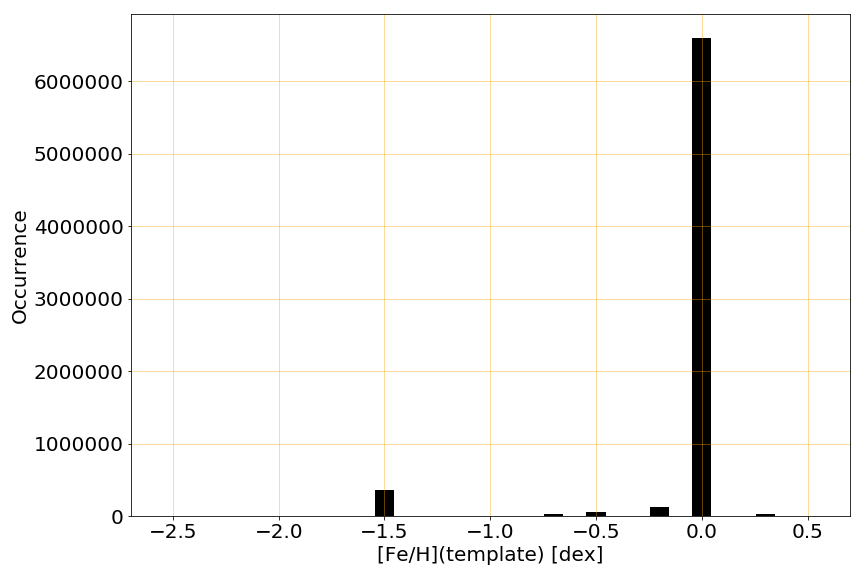}      
\caption{Distribution of the metallicities of the templates.}
\label{fig:tempFHisto}
\end{figure}

The transit radial velocities are derived by {\it "comparison"}\footnote{Cross-correlation or minimum distance methods (see \citet{DR2-DPACP-47}).} with a synthetic spectrum, referred to as template. The spectroscopic pipeline has two ways to select a template. If the parameters of the star are contained in the compilation of ground-based stellar parameters used by the pipeline (see \citet{DR2-DPACP-47}), they are used to select the closest template in a library of 5256 spectra. Otherwise, the template is chosen by a dedicated module, \emph{determineAP}, from a smaller set of 28 templates. Of the \starPublished{} stars, $\sim 18$\% had their templates selected using the ground-based compilation. Figures~\ref{fig:tempTHisto}, \ref{fig:tempGHisto} and \ref{fig:tempFHisto} show respectively the distributions of the effective temperatures, surface gravities and metallicities of the templates. The alternation of tall and small peaks in the effective temperatures distribution is the consequence of certain temperatures being chosen only when the stars are contained in the ground based catalogues compilations (producing small peaks). Similar selection effects are visible for surface gravity (few templates with $\log g \leq 2.5$ or equal to 4) and metallicity (templates mostly solar, with a {\it "small"} secondary peak at $[Fe/H] = -1.5$~dex, which are the metallicities in the sub-library used by the dedicated selection module).

The templates parameters are published in the second \gaia{} release in order to inform the users on the synthetic spectra used to derive the radial velocities. They are not meant to be used as an estimate of the stars atmospheric parameters for any other purpose. It should be noted that \gdrtwo{} contains effective temperature estimates, derived from $G$, $G_{\rm BP}$ and $G_{\rm RP}$ photometry, for about 160 million stars with $\Teff{} \in [3000, 10000]$~K and brighter than $G = 17$~mag \citep{DR2-DPACP-43}.

\subsection{Radial velocity accuracy\label{sec:gzp}}
The accuracy is the difference between the measured value and the true value. Unfortunately, the true values of the radial velocities of the \gdrtwo{} stars are not known. Therefore, as a proxy, five ground-based catalogues are used: CU6GB \citep{DR2-DPACP-48}, SIM~\citep{MakarovUnwin2015}, RAVE~\citep{Kunder2017}, APOGEE~\citep{Abolfathi2017} and Gaia-ESO-Survey~(GES; \citealt{Gilmore2012}). In each case, only a subset of these catalogues has been selected, made of stars showing no radial velocity variability. Our GES validation sub-sample being small, it has been used only for some of the accuracy assessments. 

The limitation of the comparison to external catalogues, is that they can, of course, also be affected by their own biases. Therefore, when a systematic difference is observed between \gdrtwo{} and a catalogue, it could come from the former, the latter or both. Using several external catalogues is meant to help (pending that they do not share some systematic(s)). It should be noted that, even if the true accuracy is difficult to assess from the comparison of different catalogues, it is of interest to know their relative differences, for e.g. combining data from these catalogues or comparing results obtained independently with them.

CU6GB \citep{DR2-DPACP-48} is a ground-based radial velocity catalogue produced by the \gaia{} DPAC. The zero-point of the radial velocities of the CU6GB catalogue (i.e. its accuracy) has been assessed by comparing measured radial velocities of asteroids to celestial mechanics predictions. The zero-point of CUGB is $38\pm5$~\ms. 

As \emph{estimator of the accuracy}, we use the median of the radial velocity residuals (i.e. Gaia~DR2 minus ground based catalogue). The lower and upper 1-$\sigma$ uncertainties on the estimate of the accuracy are calculated respectively as:
\begin{equation}
\epsilon_{acc}^{low} = \sqrt{{\pi}\over{2 N_{V_R^{res}}}} (\tilde{V}_R^{res} - Per(V_R^{res}, 15.85))
\end{equation}
\begin{equation}
\epsilon_{acc}^{upp} = \sqrt{{\pi}\over{2 N_{V_R^{res}}}} (Per(V_R^{res}, 84.15) -\tilde{V}_R^{res})
\end{equation}
where $N_{V_R^{res}}$ is the number of radial velocity residuals, $\tilde{V}_R^{res}$ the median of the radial velocity residuals and $Per(V_R^{res}, 15.85)$ and $Per(V_R^{res}, 84.15)$ respectively the 15.85$^{th}$ and 84.15$^{th}$ percentiles of the distribution of radial velocity residuals. In the following sections, the accuracy is studied as a function of different parameters (e.g. \extgrvs, \intgrvs, $\Teff$, $\logg$). The comparison samples are therefore divided by bins of the considered quantities. In these cases, a minimum of 20 stars per bin is required to calculate the median of the radial velocity residuals.

\begin{table}[h!]
\begin{center}
\caption[]{Median radial velocity residuals derived from the comparison of \gdrtwo{} with ground-based catalogues. N$_{stars}$ (Col. 3) is the number stars selected to calculate the median residuals.\label{tab:acc}}
\begin{tabular}{| l | c | c |} \hline
Catalogue & Median $V_R^{res}$                  & N$_{stars}$ \\
          & [$\kms$]                            &             \\ \hline
CU6GB     & $-$0.012 {\small $-$0.005/$+$0.005} & 4083        \\
SIM       & $+$0.264 {\small $-$0.015/$+$0.014} &  640        \\
RAVE      & $+$0.295 {\small $-$0.013/$+$0.013} & 9127        \\
APOGEE    & $+$0.233 {\small $-$0.011/$+$0.013} & 8124        \\ 
GES       & $+$0.236 {\small $-$0.043/$+$0.049} & 2120        \\ \hline
\end{tabular}
\end{center}
\end{table}

Table~\ref{tab:acc} presents the median radial velocity residuals derived for the five ground-based catalogues. Gaia-DR2 shows a small 2.4-$\sigma$ level offset of $-12$ $\ms$ with respect to CU6GB. The CU6GB stars brighter than or equal to \extgrvs$=9$~mag, and showing no variability, were used in the wavelength calibration procedure to set the wavelength zero-point of each configuration: i.e. field-of-view/CCD/trending epoch \citep{DR2-DPACP-47}. It is a good sanity check for the pipeline, that neither the wavelength calibration, nor the subsequent modules did modify significantly the global radial velocity zero-point defined by the CU6GB catalogue. Gaia-DR2 shows a positive global offset of about $+$200 to $+$300 $\ms$ with respect to the SIM, RAVE, APOGEE and GES stars. Yet, as discussed in the following sections, the origin these offsets varies partially from one catalogue to another.

\subsubsection{Accuracy versus number of transits}
\begin{figure}[h!]
\centering
\includegraphics[width=0.5\textwidth]{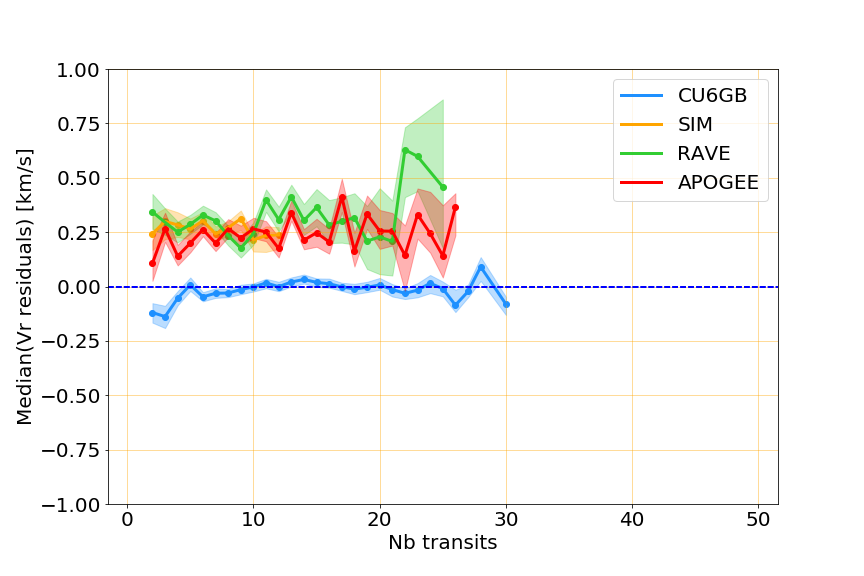}      
\caption{Median radial velocity residuals as a function of the number of transits. The lower and upper 1-$\sigma$ uncertainties on the measures of the medians are represented as shaded areas.}
\label{fig:accuNTr}
\end{figure}

Figure~\ref{fig:accuNTr} shows the median radial velocity residuals as a function of the number of eligible transits. The four accuracy curves show little trend with the number of transits. In general, the accuracy is not expected to improve or vary with the number of measures. Yet, in \gdrtwo{}, the number of transits is very correlated with the location of the stars on the sky (Figure~\ref{fig:transit}). Therefore, potential spatial biases could have been propagated to the number of transits. Figure~\ref{fig:accuNTr} indicates that this does not seems to be the case.

\subsubsection{Accurcay versus $G_\mathrm{RVS}$ \label{sec:accugrvs}}
\begin{figure}[h!]
\centering
\includegraphics[width=0.5\textwidth]{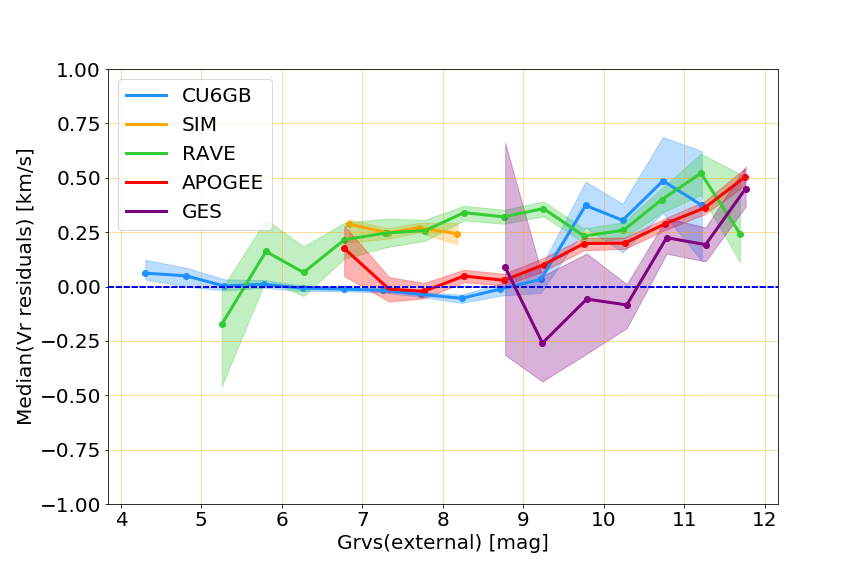}      
\caption{Median radial velocity residuals as a function of the external $G_\mathrm{RVS}^{ext}$ magnitude. The lower and upper 1-$\sigma$ uncertainties on the measures of the medians are represented as shaded areas.}
\label{fig:accuGrvs}
\end{figure}

\begin{figure}[h!]
\centering
\includegraphics[width=0.5\textwidth]{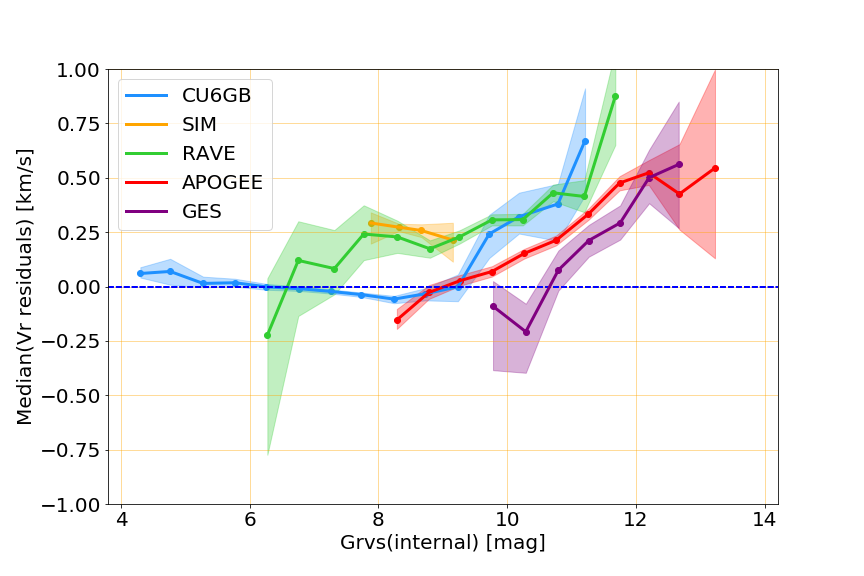}      
\caption{Median radial velocity residuals as a function of the internal $G_\mathrm{RVS}^{int}$ magnitude. The lower and upper 1-$\sigma$ uncertainties on the measures of the medians are represented as shaded areas.}
\label{fig:accuIntGrvs}
\end{figure}

Figure~\ref{fig:accuGrvs} shows the median radial velocity residuals as a function of the {\it external} $G_\mathrm{RVS}^{ext}$ magnitude. Over the magnitude range [7, 9]~mag, Gaia~DR2 shows no significant offset nor trend with respect to the CU6GB or APOGEE stars and it shows a nearly constant offset (similar to the global offset reported in Sect.~\ref{sec:gzp}) with respect to SIM  and RAVE stars. For fainter stars, beyond Grvs$\sim$9-10~mag, the \gdrtwo{} velocities exhibit an increasing positive offset with respect to all the catalogues, reaching about $\sim$500 $\ms$ at Grvs$\sim$11.75. The trend is also visible (Figure~\ref{fig:accuIntGrvs}) in the median of the residuals as a function of the {\it internal} $G_\mathrm{RVS}^{int}$ magnitude, which is derived from the flux recorded in the RVS windows.

\begin{figure}[h!]
\centering
\includegraphics[width=0.5\textwidth]{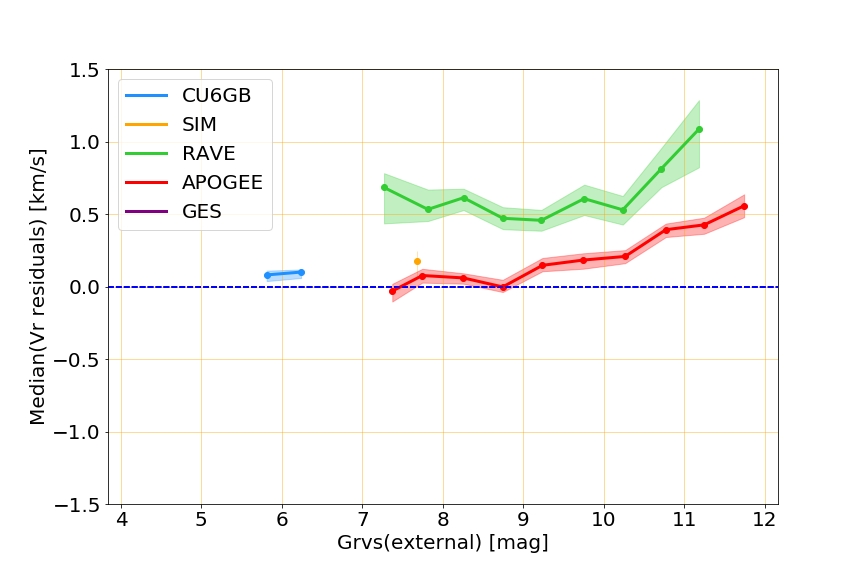}      
\caption{Median radial velocity residuals as a function of the internal $G_\mathrm{RVS}^{int}$ magnitude, for a subsets of giant stars. The lower and upper 1-$\sigma$ uncertainties on the measures of the medians are represented as shaded areas.}
\label{fig:accuGrvsGiants}
\end{figure}

Each validation catalogue contains a mix of stars of different temperatures, gravities and metallicities, whose proportion could change with magnitude. In order to check that the observed trend at faint magnitude was not caused by e.g. a differential effect between dwarfs and giants whose proportion could be function of magnitude, a sub-sample of 3220 solar metallicity giant stars was selected, fulfilling the criteria: Teff in $[4500, 5000]$~K, $\log g$ in $[2.3, 3.0]$ and $[Fe/H]$ in $[-0.3, 0.3]$~dex. Figure~\ref{fig:accuGrvsGiants} presents the median radial velocity residuals as a function of the external $G_\mathrm{RVS}^{ext}$ for the giant stars sub-sample. As for the full validation sample, at magnitude brighter than $G_\mathrm{RVS}^{ext} \sim 9$ (APOGEE) or 10~mag (RAVE) the offset is constant. At fainter magnitude the median residuals exhibit a positive gradient with magnitude.

The same trend is observed with all the validation catalogues, for both the external and internal $G_\mathrm{RVS}$ magnitudes and for the full sample as well as for a subset of similar giant stars. The probability is therefore high, that the effect is in the \gdrtwo{} data. The origin of the trend is under investigation. One of the lead is the Charge Transfer Inefficiency (CTI). When they hit the CCDs, high energy particles damage the pixels, producing traps which could snare a fraction of the spectrum photo-electron, preventing them to be propagated consistently with the rest of the signal from CDD column to CCD column. Eventually, the trapped photo-electrons are released. If the release time is short (i.e. the time for the spectrum to be propagated by one or a few pixels), the CTI does not remove signal from the spectrum, but instead distort the Line Spread Function (LSF) profile, producing a trail in the direction opposite to the propagation of the spectrum. In the RVS, the blue edge of the spectrum is leading. The CTI would therefore produce a tail on the LSF red edge, shifting the line centroids to higher wavelengths and applying a positive shift to the radial velocities. Pre-launch ground-based laboratory tests have shown that the impact of CTI increases as the signal decreases. We would therefore expect from CTI a positive radial velocity trend with magnitude, which is what we observe. Tests are planned to check if other predictions of the CTI model are fulfilled by the data (e.g. increase of the effect with time, modulation of the radial velocity shift with background signal level), aiming to understand, model and correct the effect in \gdrthree{}.

It should be noted that while SIM and RAVE validation samples present already an offset for bright stars (which increases further for RAVE faint stars), APOGEE bright stars present little offset. The  global offset between \gdrtwo{} and APOGEE reported in Sect.~\ref{sec:gzp} is mainly due to the prevalence of faint stars in this validation sample, while there is a global offset with respect to RAVE and SIM stars. Table~\ref{tab:acc9} presents the median radial velocity residuals for the subsets of stars, from the validation catalogues, brighter than $G_\mathrm{RVS}^{ext} = 9$~mag.

\begin{table}[h!]
\begin{center}
\caption[]{Median radial velocity residuals derived from the comparison of \gdrtwo{} with stars from the ground-based catalogues brighter than $G_\mathrm{RVS}^{ext} = 9$~mag. N$_{stars}$ (Col. 3) is the number stars selected to calculate the median residuals.\label{tab:acc9}}
\begin{tabular}{| l | c | c |} \hline
Catalogue & Median $V_R^{res}$                  & N$_{stars}$ \\
          & [$\kms$]                            &             \\ \hline
CU6GB     & $-$0.017 {\small $-$0.005/$+$0.005} & 3805        \\
SIM       & $+$0.264 {\small $-$0.015/$+$0.014} &  640        \\
RAVE      & $+$0.282 {\small $-$0.019/$+$0.019} & 3838        \\
APOGEE    & $+$0.024 {\small $-$0.015/$+$0.017} & 1141        \\ 
GES       & $-$0.184 {\small $-$0.418/$+$0.254} &   63        \\ \hline
\end{tabular}
\end{center}
\end{table}

\begin{figure}[h!]
\centering
\includegraphics[width=0.5\textwidth]{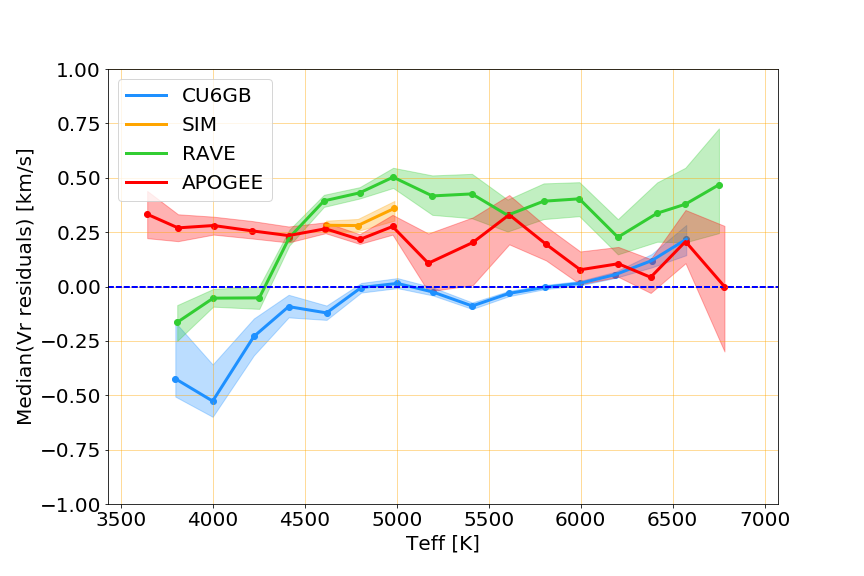}
\includegraphics[width=0.5\textwidth]{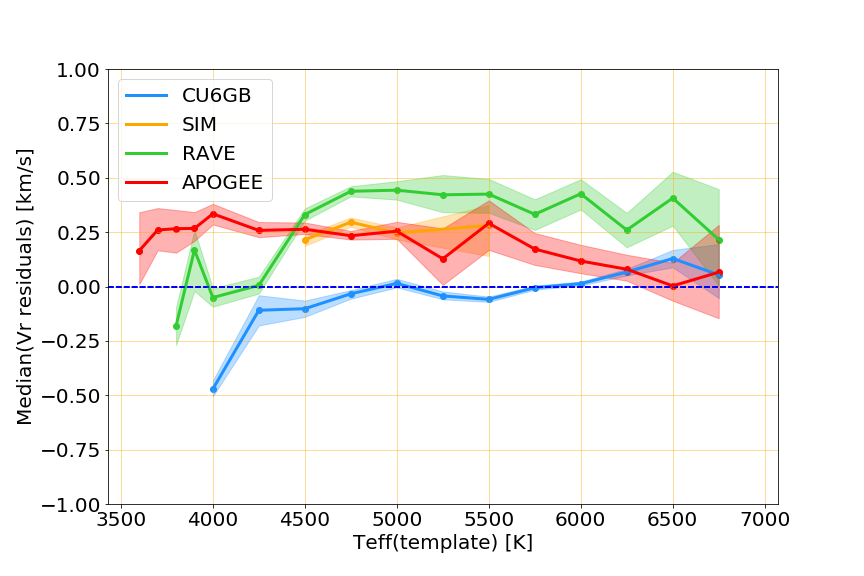}
\caption{Median radial velocity residuals as a function of the effective temperature of the stars (top) and of the templates (bottom). The lower and upper 1-$\sigma$ uncertainties on the measures of the medians are represented as shaded areas.}
\label{fig:accuTeff}
\end{figure}

\subsubsection{Accuracy versus temperature}
Figure~\ref{fig:accuTeff} presents the median radial velocity residuals as a function of the stars effective temperature (top) and as a function of the templates effective temperature (bottom). Both the CU6GB and the RAVE validation samples show a drop of the median residuals by about 500~$\ms$ for cool stars, from 4750-4500 to 4000~K. Over the same temperature range, the APOGEE sample exhibits flat residuals. Beyond, $\sim$5500~K the median residuals of the APOGEE sample show a smooth decrease of about 200~$\ms$.

\subsubsection{Accuracy versus gravity}
\begin{figure}[h!]
\centering
\includegraphics[width=0.5\textwidth]{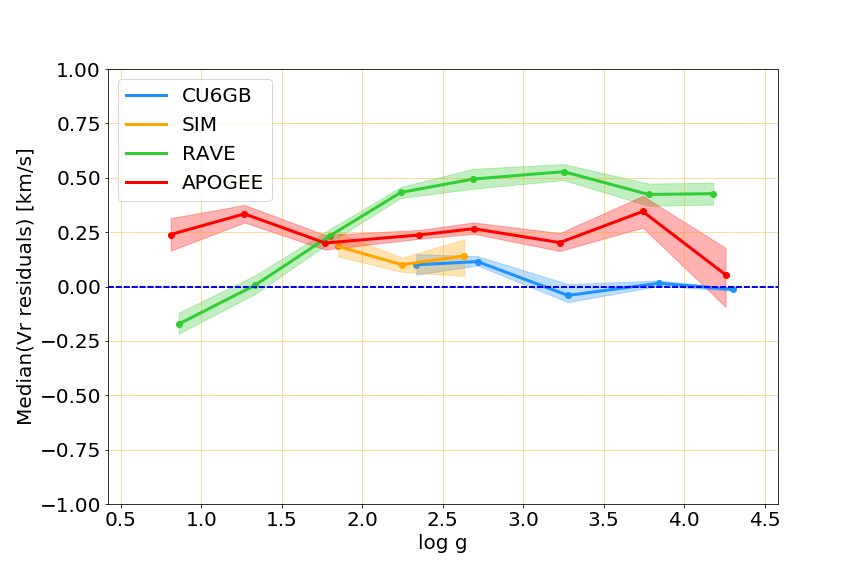}
\includegraphics[width=0.5\textwidth]{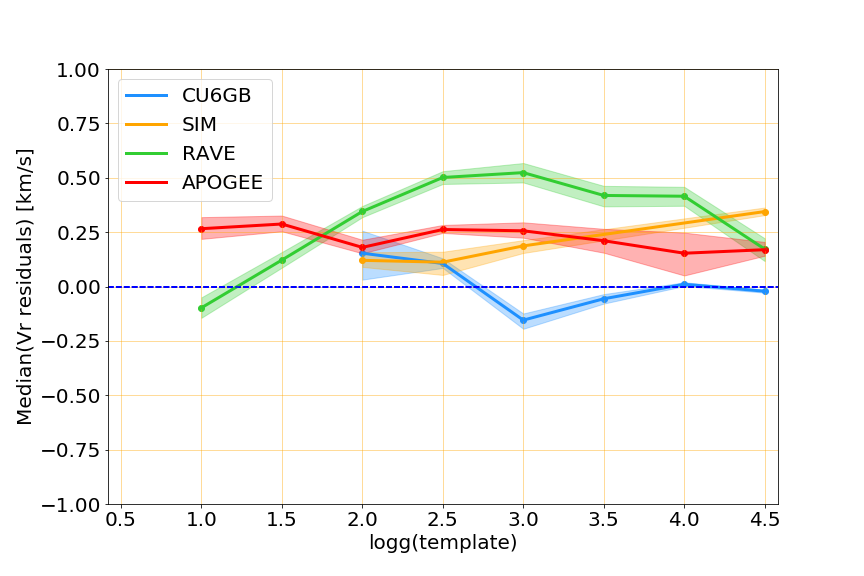}
\caption{Median radial velocity residuals as a function of the surface gravity of the stars (top) and of the templates (bottom). The lower and upper 1-$\sigma$ uncertainties on the measures of the medians are represented as shaded areas.}
\label{fig:accuLogg}
\end{figure}

Figure~\ref{fig:accuLogg} presents the median radial velocity residuals as a function of the stars surface gravity (top) and as a function of the templates surface gravity (bottom). The main feature is a smooth decrease of the median residuals of the \gdrtwo{} data versus RAVE by about 700~$\ms$ between $\log g \sim 2.5$ and $\log g \sim 1$. The effect is not seen in the other validation samples, in particular APOGEE which also extends to low gravities.  

\subsubsection{Accuracy versus metallicity}
\begin{figure}[h!]
\centering
\includegraphics[width=0.5\textwidth]{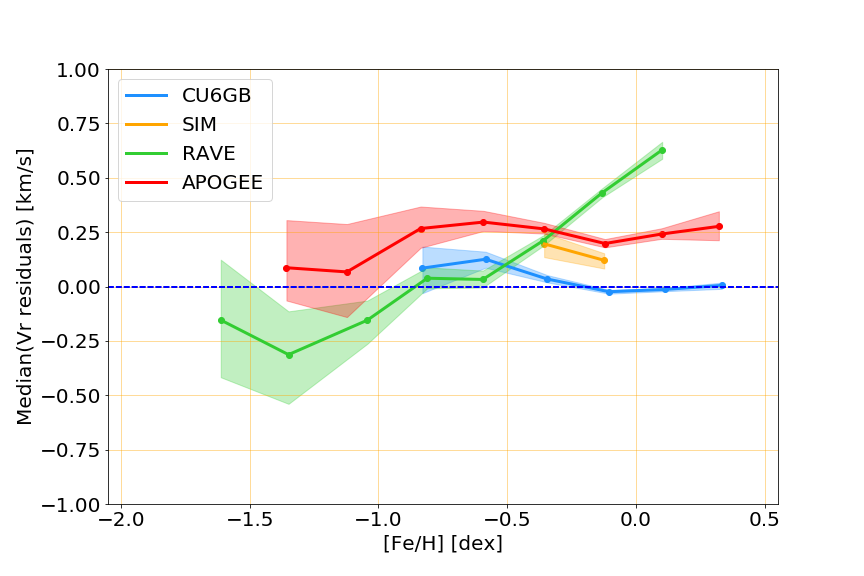}
\includegraphics[width=0.5\textwidth]{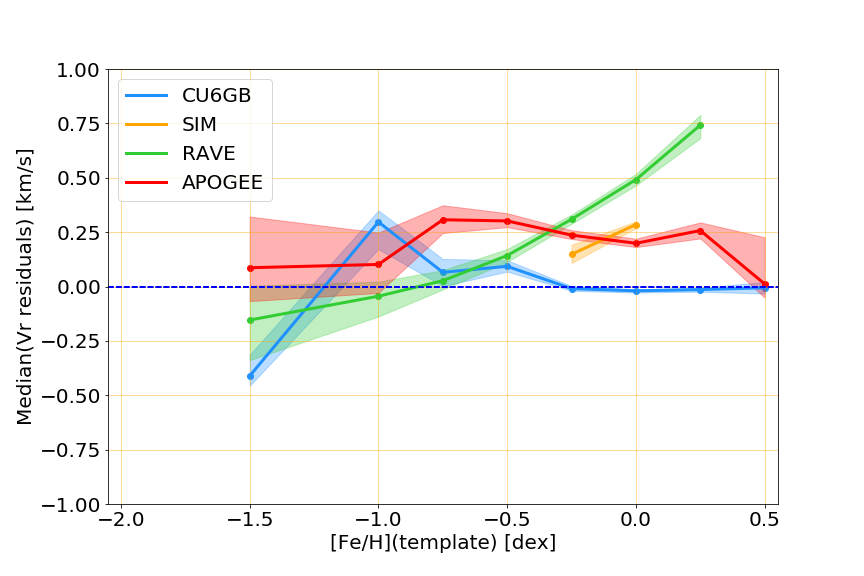}
\caption{Median radial velocity residuals as a function of the metallicity of the stars (top) and of the templates (bottom). The lower and upper 1-$\sigma$ uncertainties on the measures of the medians are represented as shaded areas.}
\label{fig:accuFeH}
\end{figure}

Figure~\ref{fig:accuFeH} presents the median radial velocity residuals as a function of the stars metallicity (top) and as a function of the templates metallicity (bottom). The median residuals of the \gdrtwo{} versus RAVE velocities show a positive trend with metallicity which reaches $\sim$750~$\ms$ for the most metal-rich sources. On the metal-poor side, the CU6GB sample shows a negative offset of about $-500$~$\ms$ for the stars whose velocity has been calculated with templates having a metallicity of $-1.5$~dex. The APOGEE validation sample shows no significant trend with metallicity. 

\subsubsection{Accuracy versus velocity}
\begin{figure}[h!]
\centering
\includegraphics[width=0.5\textwidth]{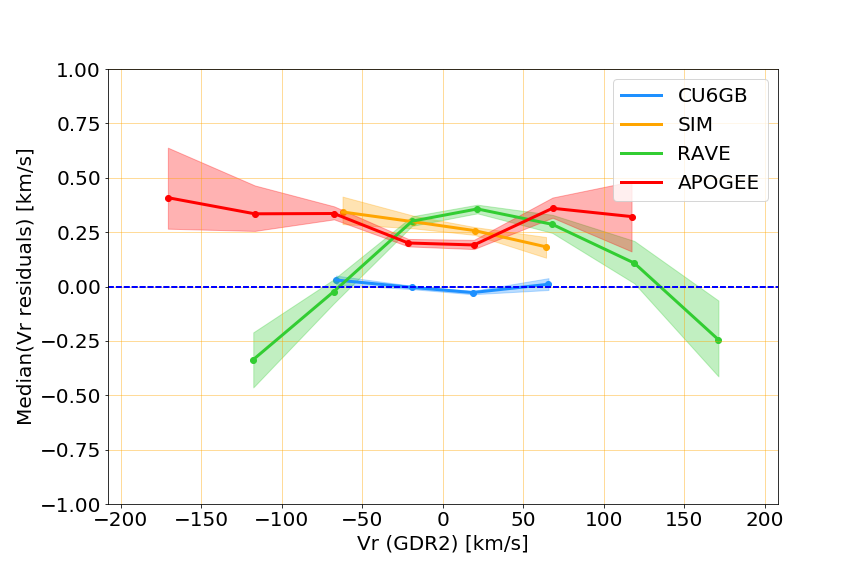}
\caption{Median radial velocity residuals as a function of the \gdrtwo{} radial velocity. The lower and upper 1-$\sigma$ uncertainties on the measures of the medians are represented as shaded areas.}
\label{fig:accuVrSpe}
\end{figure}

\begin{figure}[h!]
\centering
\includegraphics[width=0.5\textwidth]{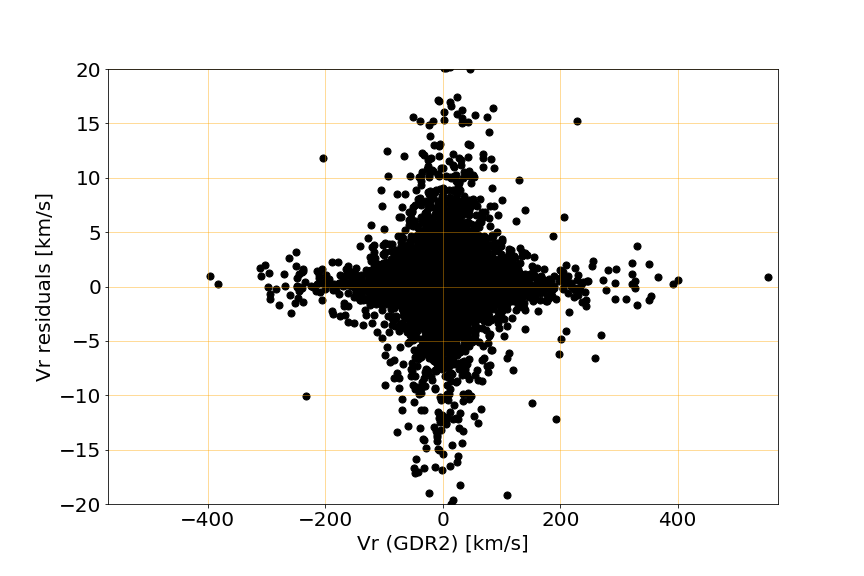}
\caption{Individual radial velocity residuals as a function of the \gdrtwo{} radial velocity, for 24~063 validation stars. Thirty one validation stars, with an absolute value of the radial velocity residual larger than 20~$\kms$, are not displayed. The dots extend to larger residuals at small absolute velocities than at larger ones. This is a visual effect due to the fact that small absolute velocities are much more densely populated than larger ones and therefore the wings of the distribution of residuals are probed to much larger values.}
\label{fig:resVrSpe}
\end{figure}

Figure~\ref{fig:accuVrSpe} presents the median radial velocity residuals as a function of the stars velocity. The residuals of the \gdrtwo{} versus RAVE velocities decrease by about 600-700~$\ms$ from $+$25~$\kms$ to $-$125~$\kms$ and show a symmetric behaviour at positive velocities. The small number of high velocity stars in our validation datasets prevents from deriving precise median residuals outside $[-175, +175]$~$\kms$, whereas the pipeline derives radial velocities in the range $[-1000, +1000]$~$\kms$. The individual radial velocity residuals (Figure~\ref{fig:resVrSpe}) provide a view on a broader velocity intervals. The individual residuals do not show any strong offset or trend over the interval $[-400, 400]$~$\kms$. The fastest validation star has a radial velocity of 552.6~$\kms$ in our RAVE list and 553.5~$\kms$ in \gdrtwo{}. As described in Sect.~\ref{sec:filters}, the combined spectra of the stars with radial velocities $|V_R| \geq 500$~$\kms$ were visually inspected one by one and those considered as \emph{false high-velocity stars} were discarded from \gdrtwo{}.

\subsubsection{Accuracy versus sky coordinates\label{sec:accuSky}}
\begin{figure}[h!]
\centering
\includegraphics[width=0.5\textwidth]{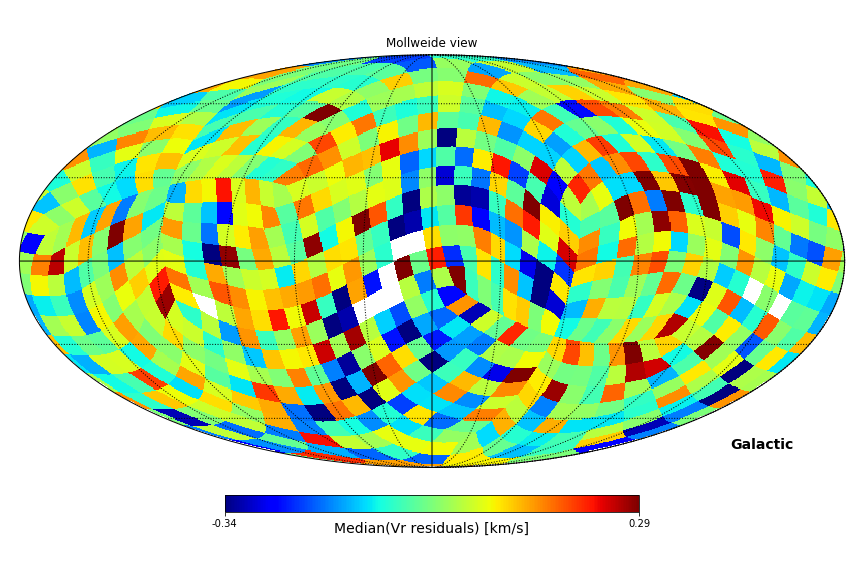}
\caption{Sky map, in galactic coordinates, of the median radial velocity residuals per pixel of $\sim$54~square degrees (healpix level 3). The Galactic Centre is in the middle of the figure and the galactic longitudes increase to the left.}
\label{fig:mapVrAccu}
\end{figure}

Figure~\ref{fig:mapVrAccu} presents the sky map, in galactic coordinates, of the median radial velocity residuals per pixel of $\sim$54~square degrees. To increase the number of validation stars, the 5 ground-based catalogues presented in Sect.~\ref{sec:gzp} have been complemented with 5820 stars from the Extended Hipparcos Compilation (XHIP; \citealt{AndersonFrancis2012}). Nonetheless, for a few pixels, the minimum number stars required to calculate the median, lowered here to 5 stars per pixel, was not reach. Those pixels appear in white on the map. To produce the map, the validation datasets have been corrected for their respective median radial velocity residuals (see Table~\ref{tab:acc}) and then combined. This explains why the distribution of median radial velocity residuals is roughly centred on zero. The 2$^{nd}$ and 98$^{th}$ percentiles of the distribution are respectively $-$0.34 and $+$0.29~$\kms$, while the extrema are $-$0.81 and $+$0.64~$\kms$.

\subsubsection{Accuracy: summary}
The main systematic found in the \gdrtwo{} radial velocities is a trend with magnitude, which start around Grvs$\sim$9-10~mag and reaches about $+$500~$\ms{}$ at \extgrvs~$= 11.75$~mag (see Sect.~\ref{sec:accugrvs}). In addition to the this trend, \gdrtwo{} shows offsets of about $+$250/$+$300~$\ms{}$ with respect to the SIM and RAVE validation samples. Other offsets, specific to a range of parameter and to a catalogue have been identified (see previous sections for the details). They do not exceed a few 100s~$\ms{}$.

\subsection{Radial velocity precision}
Two different datasets and statistical estimators are used to assess the precision of the \gdrtwo{} radial velocities.

The first dataset is made of a compilation of 6 ground-based catalogues, i.e. the five used to assess the accuracy of \gdrtwo{} radial velocities\footnote{CU6GB, SIM, RAVE, APOGEE and GES (Sect.~\ref{sec:gzp}).} plus 5820 XHIP stars~\citep{AndersonFrancis2012}. Since the different catalogues have small relative offsets, they were first corrected for their median radial velocity residuals (as listed in Table~\ref{tab:acc}) before being combined in a single dataset, hereafter referred to as GB\footnote{standing for ground-based.} validation dataset. With this dataset, the \gdrtwo{} radial velocity precision is calculated as the robust dispersion of the radial velocity residuals:
\begin{equation}
\sigma_{V_R}^{GB} = {Per(V_{R}^{res}, 84.15) - Per(V_{R}^{res}, 15.85) \over 2}
\label{eq:precResid}
\end{equation}
where $Per(V_{R}^{res}, 15.85)$ and $Per(V_{R}^{res}, 84.15)$ are respectively the 15.85$^{th}$ and 84.15$^{th}$ percentiles of the distribution of radial velocity residuals: $V_{R}^{res} = V_{R}^{GDR2} - V_{R}^{GB}$.
The lower and upper 1-$\sigma$ uncertainties on the precision are calculated as:
\begin{equation}
\epsilon_{\sigma_{V_R}^{GB}}^{low} = {\sqrt{2 \pi} \over e^{-0.5}} \sqrt{0.1585 \times 0.683 \over N_{V_R^{res}}} (\tilde{V}_R^{res} - Per(V_R^{res}, 15.85))
\end{equation}
\begin{equation}
\epsilon_{\sigma_{V_R}^{GB}}^{upp} = {\sqrt{2 \pi} \over e^{-0.5}} \sqrt{0.1585 \times 0.683 \over N_{V_R^{res}}} (Per(V_R^{res}, 84.15) - \tilde{V}_R^{res})
\end{equation}
where $\tilde{V}_R^{res}$ is the median of the radial velocity residuals and $N_{V_R^{res}}$ the number of radial velocity residuals.

The second dataset is made of all the stars with a radial velocity published in \gdrtwo{} and is hereafter referred to as the full dataset. With this dataset, the estimator of the precision is the median of the radial velocity uncertainties (Sect.~\ref{sec:vrError}):
\begin{equation}
\sigma_{V_R}^{Full} = \tilde{\epsilon}_{V_R}
\label{eq:precUncer}
\end{equation}
and the lower and upper 1-$\sigma$ uncertainties on the precision are calculated as:
\begin{equation}
\epsilon_{\sigma_{V_R}^{Full}}^{low} = \sqrt{{\pi}\over{2 N_{\epsilon_{V_R}}}} (\tilde{\epsilon}_{V_R} - Per(\epsilon_{V_R}, 15.85))
\end{equation}
\begin{equation}
\epsilon_{\sigma_{V_R}^{Full}}^{upp} = \sqrt{{\pi}\over{2 N_{\epsilon_{V_R}}}} (Per(\epsilon_{V_R}, 84.15) - \tilde{\epsilon}_{V_R})
\end{equation}
where $N_{\epsilon_{V_R}}$ is the number of individual radial velocity uncertainties and $\tilde{\epsilon}_{V_R}$, $Per(\epsilon_{V_R}, 15.85)$ and $Per(\epsilon_{V_R}, 84.15)$ respectively the median, 15.85$^{th}$ and 84.15$^{th}$ percentiles of the distribution of radial velocity uncertainties. The radial velocity uncertainty, $\epsilon_{V_R}$, is function of the standard deviation of the time series of transit radial velocities. The precision derived from the full validation dataset is therefore proportional to the scatter of the transit radial velocities.

Although the full dataset is made of the 7.2 million stars, the effective temperature, surface gravity and metallicity are known respectively for only 1.3, 0.4 and 0.4 million stars (from a compilation of ground-based catalogues). The assessment of the radial velocity precision as a function of these parameters is therefore limited to these sub-samples.

\begin{figure}[h!]
\centering
\includegraphics[width=0.45\textwidth]{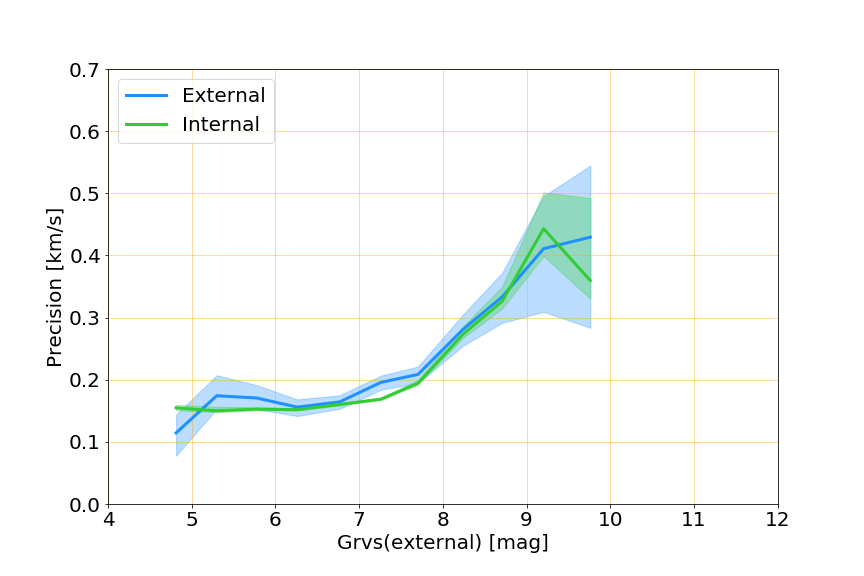}
\caption{Comparison of the external and internal precisions, respectively estimated using Eq.~\ref{eq:precResid} and Eq.~\ref{eq:precUncer}, as a function of \extgrvs{} magnitude, for the CU6GB validation stars.}
\label{fig:precisions}
\end{figure}

To compare the estimates of the precision derived with the robust dispersion of the residuals (Eq.~\ref{eq:precResid}) on the one hand and with the median of the radial velocity uncertainties (Eq.~\ref{eq:precUncer}) on the other hand, both estimators were applied to the same subset of CU6GB stars. Figure~\ref{fig:precisions} shows the \emph{external} (relying on the residuals) and \emph{internal} (relying on the uncertainties) precisions as a function of \extgrvs{} magnitude. Within the uncertainties, the two estimators of the precision are in good agreement.

The two validation datasets and the two estimators of the precision are complementary, in the sense where they are sensitive to different aspects. The stars in the GB dataset have been selected on the basis of multiple ground-based observations to be as clean as possible of radial velocity variables. The full dataset contains all the \starPublished{} stars of the \gdrtwo{} spectroscopic catalogue, which includes a proportion of undetected multiple and variable stars, even if the filters applied on the suspected SB2 and on the radial velocity uncertainty (Sect.~\ref{sec:filters}) should have removed the bulk of the large amplitude velocity variables. Furthermore, the GB based precision includes a contribution from the ground-based catalogues themselves. For the ground-based velocities derived from high-resolving power echelle spectrograph (like in the CU6GB catalogue), this contribution is very small. The GB precision will account for some potential differential effects, like e.g. possible systematic offset between giant and dwarf stars. The full dataset precision, which relies on the scatter of the transit radial velocities of each source, is not sensitive to the possible systematics between different groups of sources.

\begin{figure}[h!]
\centering
\includegraphics[width=0.45\textwidth]{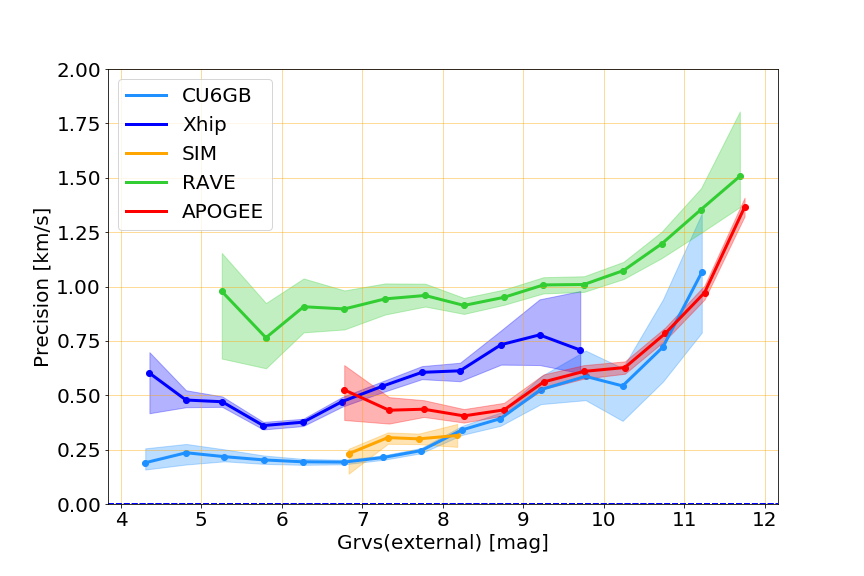}
\caption{Radial velocity precision as a function of \extgrvs{} magnitude, for the ground-based catalogues composing the GB validation dataset. The lower and upper 1-$\sigma$ uncertainties on the measures of the precision are represented as shaded areas.}
\label{fig:precGB}
\end{figure}

\begin{figure}[h!]
\centering
\includegraphics[width=0.40\textwidth]{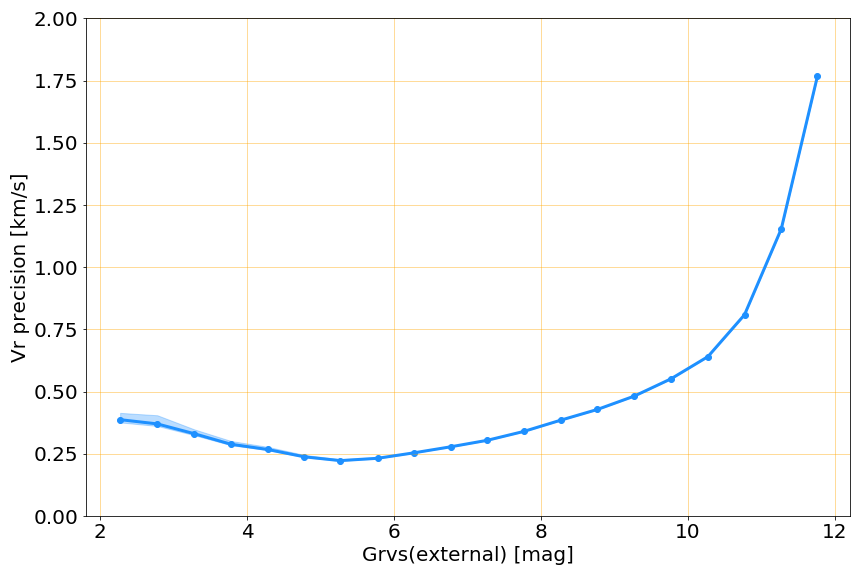}
\caption{Radial velocity precision as a function of \extgrvs{} magnitude, assessed using the full sample. The lower and upper 1-$\sigma$ uncertainties on the measures of the precision are represented as shaded areas.}
\label{fig:precAll}
\end{figure}

\subsubsection{Precision versus magnitude}
Figure~\ref{fig:precGB} shows the precision, estimated from the robust dispersion of the radial velocity residuals (Eq.~\ref{eq:precResid}), as a function of the external \extgrvs{} magnitude, for the ground-based catalogues included in the GB validation dataset. Figure~\ref{fig:precAll} shows the precision, estimated from the median of the radial velocity uncertainties (Eq.~\ref{eq:precUncer}), as a function of the external \extgrvs{} magnitude, for the full dataset. The precision derived using the CU6GB sample reaches $\sim$200-250~$\ms$ for the stars with \extgrvs{}~$\in [4, 8]$~mag, close to the precision obtained with the full dataset for \extgrvs{}~$\in [4, 8]$~mag: $\sim 220-350$~$\ms$. This is about 3 to 5 times more precise than the pre-launch specification, which was 1~$\kms$. In this range of magnitude, the \gdrtwo{} radial velocity precision is limited by the precision of the calibrations and in particular of the wavelength calibration. From \extgrvs{}~$\sim 5$~mag, some spectra begin to be saturated. The proportion increases with decreasing magnitude and as a consequence the radial velocity precision deteriorates (Fig.~\ref{fig:precAll}). The estimation of the bright stars precision using Xhip, RAVE and APOGEE is limited by the internal uncertainties of these catalogues. Beyond,  \extgrvs{}~$\sim 9$~mag, the CU6GB and APOGEE samples show similar behaviours. At \extgrvs{}~$= 11.75$~mag, the APOGEE and RAVE stars yield similar \gdrtwo{} precisions of 1.4-1.5 $\kms$. The precision derived with the full sample is slightly lower, i.e. 1.8~$\kms$, probably because of the larger proportion of unfiltered radial velocity variables.


\begin{figure}[h!]
\centering
\includegraphics[width=0.45\textwidth]{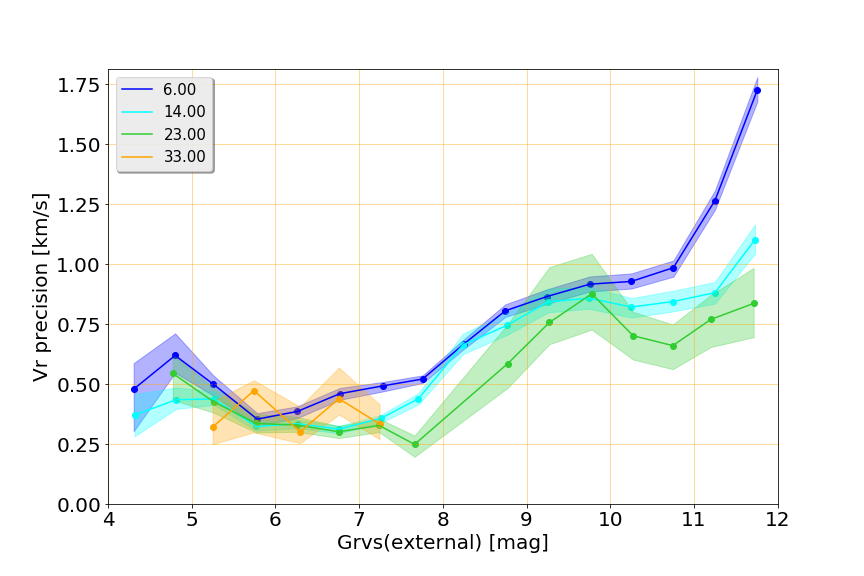}
\includegraphics[width=0.45\textwidth]{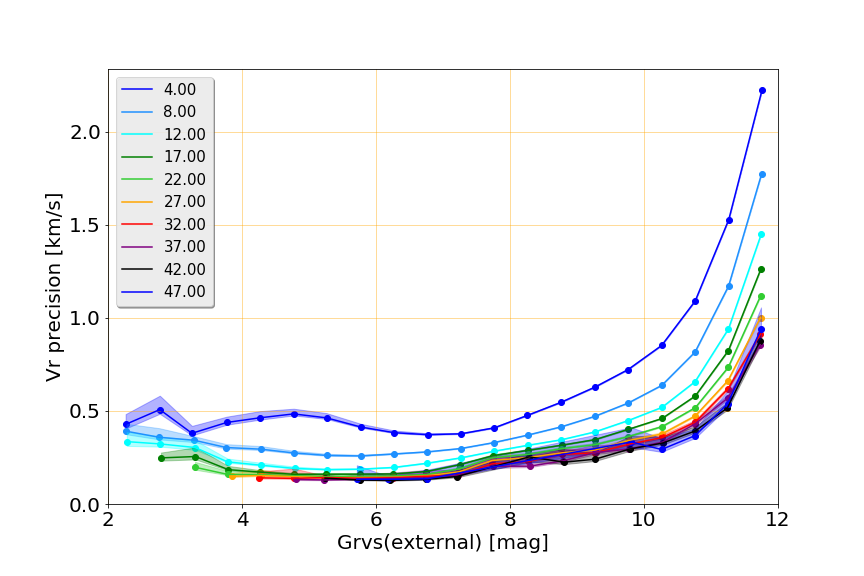}
\caption{Radial velocity precision as a function of \extgrvs{} magnitude, respectively for the GB validation dataset (top) and for the full dataset (bottom). The curves have been calculated for different ranges of number of transits. The mean values of the intervals are given in the captions. The lower and upper 1-$\sigma$ uncertainties on the measures of the precision are represented as shaded areas.}
\label{fig:precNTr}
\end{figure}

\subsubsection{Precision versus number of transits}
Figure~\ref{fig:precNTr} presents the radial velocity precision as a function of \extgrvs{} magnitude and number of transits (the different curves), respectively for the GB validation dataset (top) and for the full dataset (bottom). As expected, the precision improves with the number of transits. For example, at \extgrvs{}~$= 11.75$~mag, the precision estimated using the full dataset, improves from $\sim 2.2$ to $\sim 0.9$~$\kms$, between 4 and 37 transits.

\begin{figure}[h!]
\centering
\includegraphics[width=0.5\textwidth]{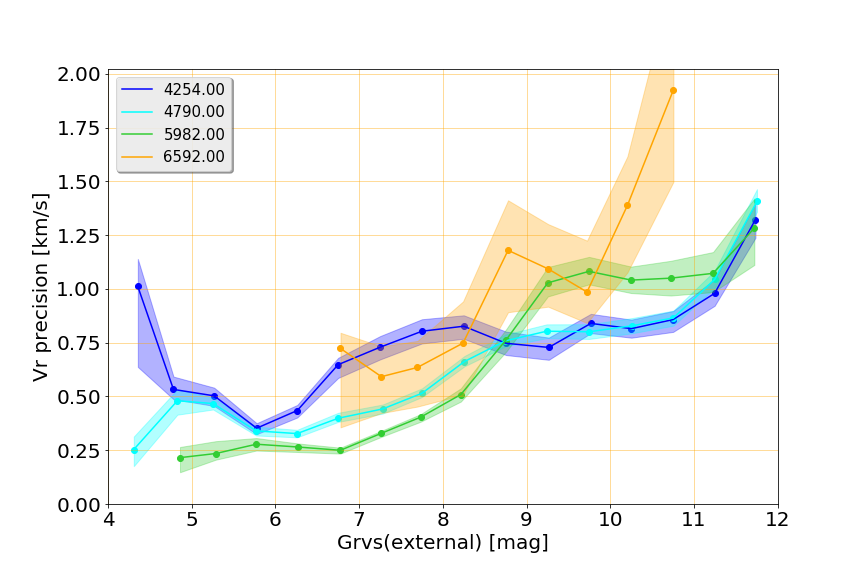}
\includegraphics[width=0.5\textwidth]{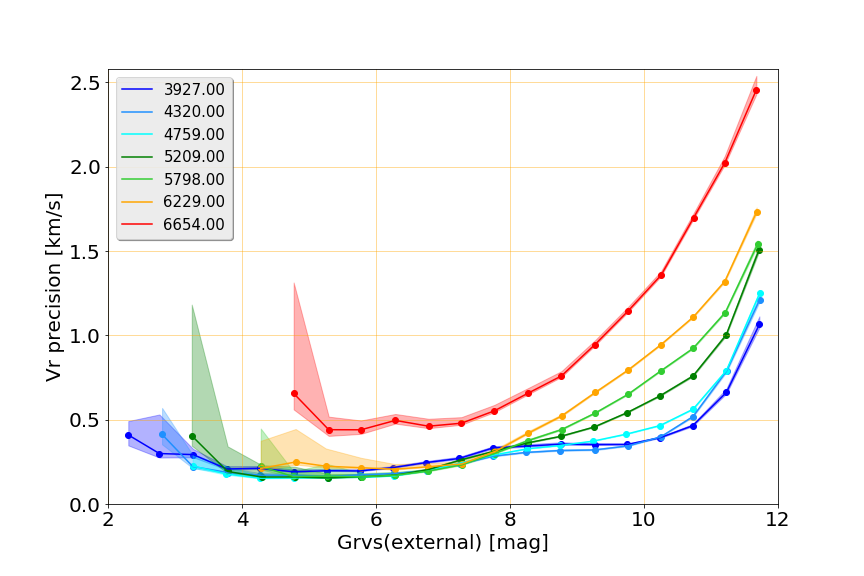}
\caption{Same as Fig.~\ref{fig:precNTr}. The curves have been calculated for different effective temperature ranges.}
\label{fig:precTeff}
\end{figure}

\subsubsection{Precision versus effective temperature\label{sec:teff}}
\begin{figure}[h!]
\centering
\includegraphics[width=0.5\textwidth]{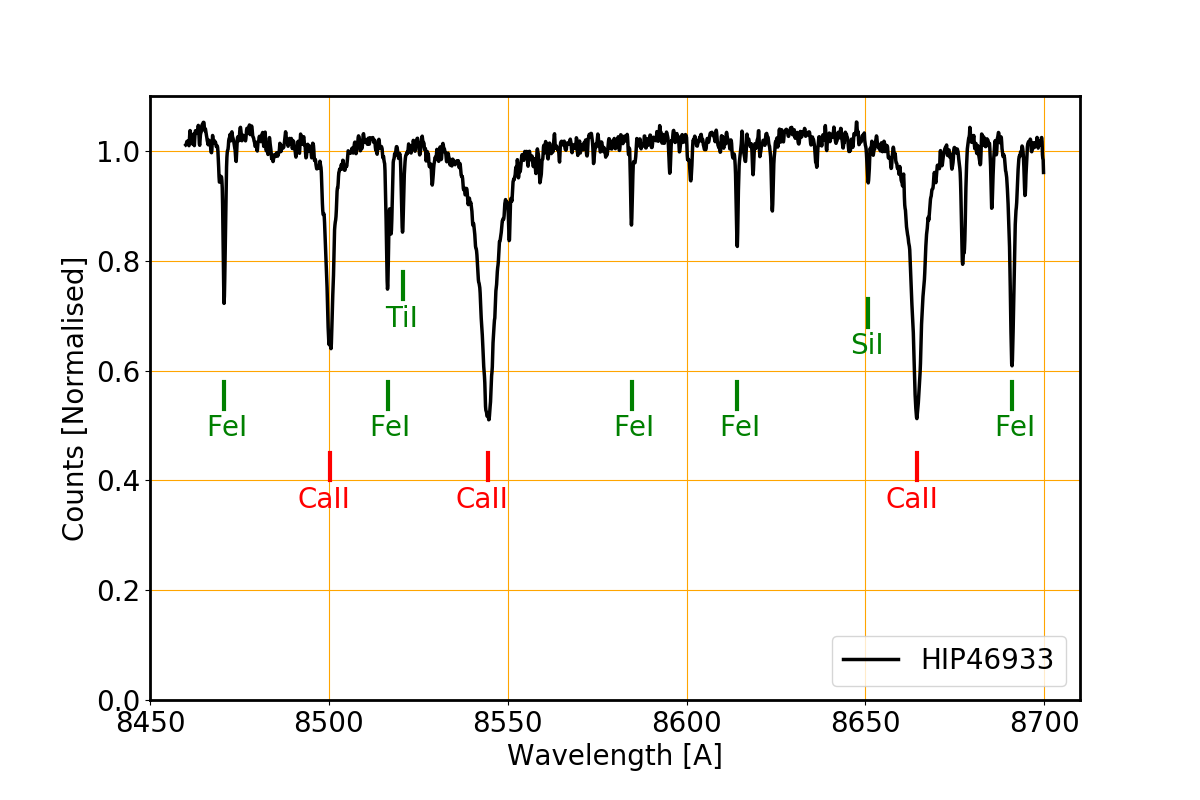}
\includegraphics[width=0.5\textwidth]{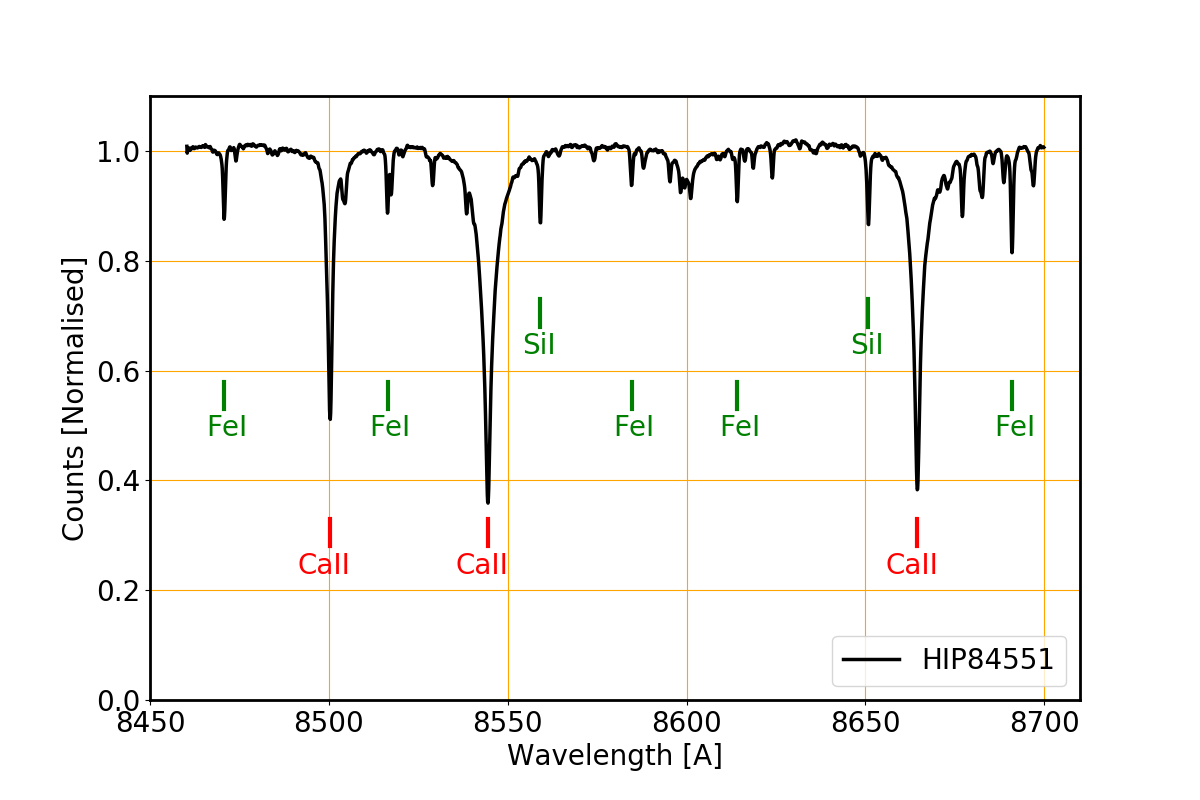}
\caption{Comparison of the RVS spectra of HIP46933 and HIP84551. The weak neutral lines are on average stronger in the former which is cooler, $\Teff = 4487$~K, than in the latter, $\Teff = 6517$~K and therefore allow to derive higher precision radial velocities (at similar \grvs{} magnitude).}
\label{fig:hip46933}
\end{figure}

Figure~\ref{fig:precTeff} presents the \gdrtwo{} precision as a function of the external \extgrvs{} magnitude and effective temperature (the different curves), respectively for the GB validation dataset (top) and for 1.3 million stars from the full dataset with known $\Teff{}$ (bottom). The radial velocity precision improves as the effective temperature decreases. This is the direct consequence of the evolution of the morphology of the spectra with temperature. As it decreases, the weak neutral lines become on average stronger, carrying more information to derive the radial velocity. This is illustrated by Fig.~\ref{fig:hip46933} which compares the spectra of a HIP46933 ($\Teff = 4487$~K, $\logg = 4.22$, $\FeH = -0.24$~dex; \citealt{Adibekyan2012}) and HIP84551 ($\Teff = 6517$~K, $\logg = 4.20$, $\FeH = +0.18$~dex; \citealt{Adibekyan2012}). At  \extgrvs{}~$= 11.75$~mag, the precision estimated from the full sample is $\sim 2.5$~$\kms$ at $\Teff \sim 6650$~K, $\sim 1.5$~$\kms$ at $\Teff \sim 5800$~K and $\sim 1.1$~$\kms$ at $\Teff \sim 3900$~K. 

\begin{figure}[h!]
\centering
\includegraphics[width=0.5\textwidth]{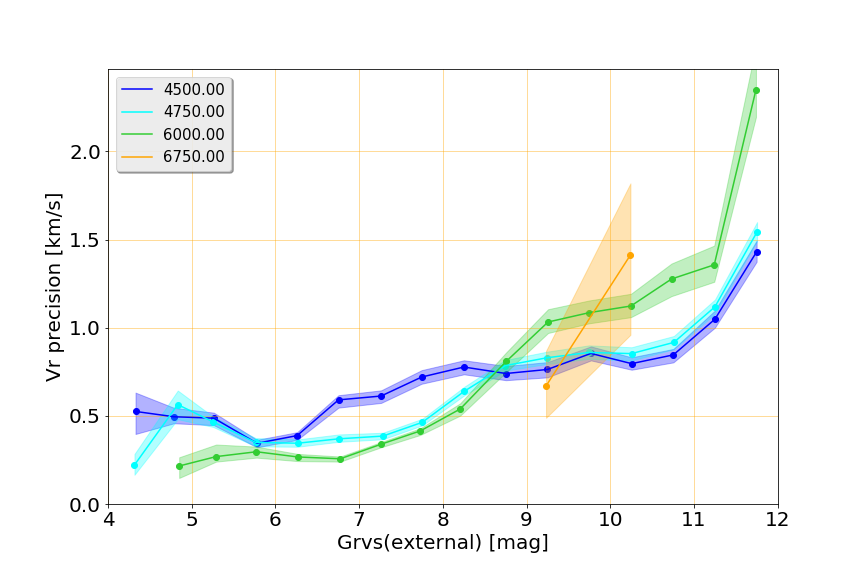}
\includegraphics[width=0.5\textwidth]{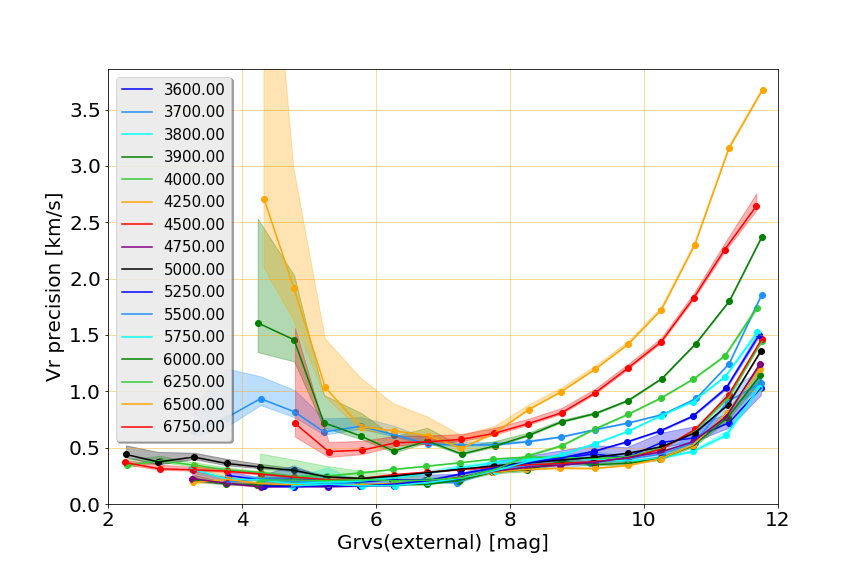}
\caption{Same as Fig.~\ref{fig:precNTr}. The curves have been calculated for different template effective temperature ranges.}
\label{fig:precTempT}
\end{figure}

Figure~\ref{fig:precTempT} is similar to Figure~\ref{fig:precTeff}, with the curves corresponding to different interval of effective temperatures of the templates used to process the stars (rather than interval of effective temperature of the stars). Globally, the precision improves as the effecive temperature of the template decreases. Locally, some adjacent curves could show the opposite behaviour. For example, with the full sample, a better precision is obtained for the stars processed with a \tempTeff{}~$= 6750$~K template (red curve), than with a \tempTeff{}~$= 6500$~K template (orange curve). The sub-library used by the software module {\it determineAP} includes a synthetic spectra with \tempTeff{}~$= 6500$~K, but none with \tempTeff{}~$= 6750$~K. Therefore, stars which have been processed with a \tempTeff{}~$= 6750$~K template all had atmospheric parameters included in the compilation of ground-based catalogues. Those processed with a \tempTeff{}~$= 6500$~K template are in majority not included in the ground-based compilation and their templates has been selected by {\it determineAP}. The stars whose templates have been selected by {\it determineAP} suffer from a larger template mismatch than those included in the ground-based compilation, which explains the swapping of some curves. At \extgrvs{}~$= 11.75$~mag, the precision estimated from the full sample is $\sim 3.7$~$\kms$ at \tempTeff{}~$\sim 6500$~K, $\sim 2.6$~$\kms$ at \tempTeff{}~$\sim 6750$~K, $\sim 1.5$~$\kms$ at \tempTeff{}~$\sim 5750$~K, $\sim 1.4$~$\kms$ at \tempTeff{}~$\sim 5000$~K and $\sim 1.1$~$\kms$ at \tempTeff{}~$\sim 3900$~K.

The precision as a function of effective temperature (Fig.~\ref{fig:precTeff}) is representative of the performance achieved for the 1.3 million stars with known ground-based effective temperature and therefore with the smallest template mismatches. The precision as a function of the template effective temperature (Fig.~\ref{fig:precTempT}) is representative of the full 7.2 million \gdrtwo{} stars and account for the full variety of template mismatches. The faint stars precisions as a function of magnitude and effective temperatures quoted in the abstract and conclusion are therefore those estimated using the full sample split by template effective temperature range, which are more representative of the \gdrtwo{} radial velocities as a whole.

\subsubsection{Precision versus surface gravity}
\begin{figure}[h!]
\centering
\includegraphics[width=0.5\textwidth]{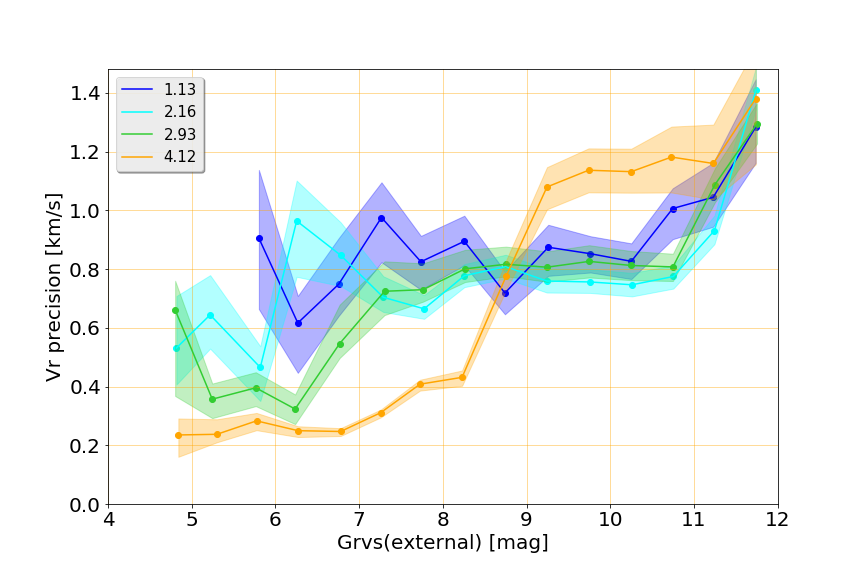}
\includegraphics[width=0.5\textwidth]{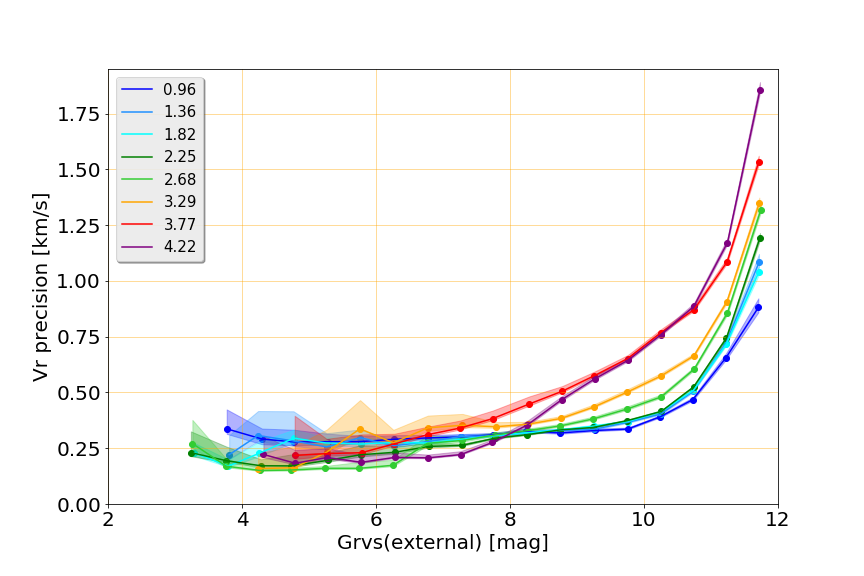}
\caption{Same as Fig.~\ref{fig:precNTr}. The curves have been calculated for different surface gravity ranges.}
\label{fig:precLogg}
\end{figure}

\begin{figure}[h!]
\centering
\includegraphics[width=0.5\textwidth]{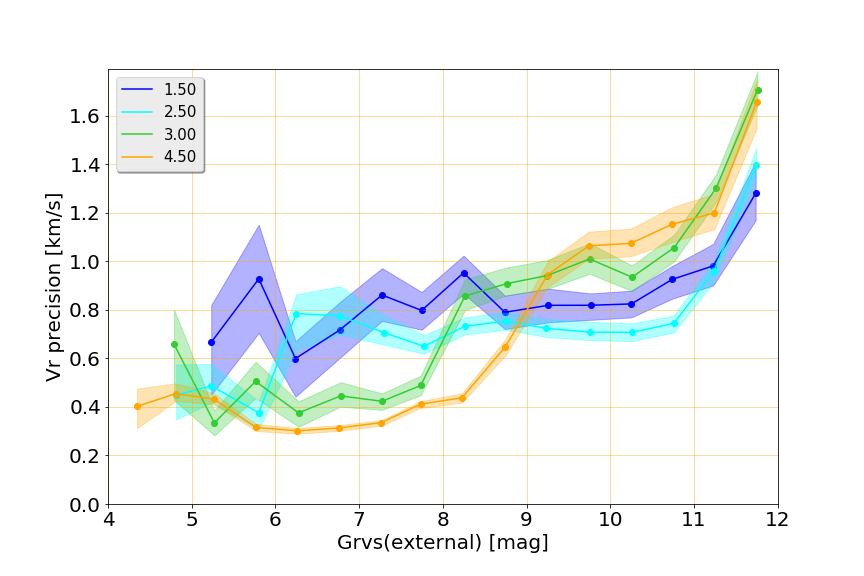}
\includegraphics[width=0.5\textwidth]{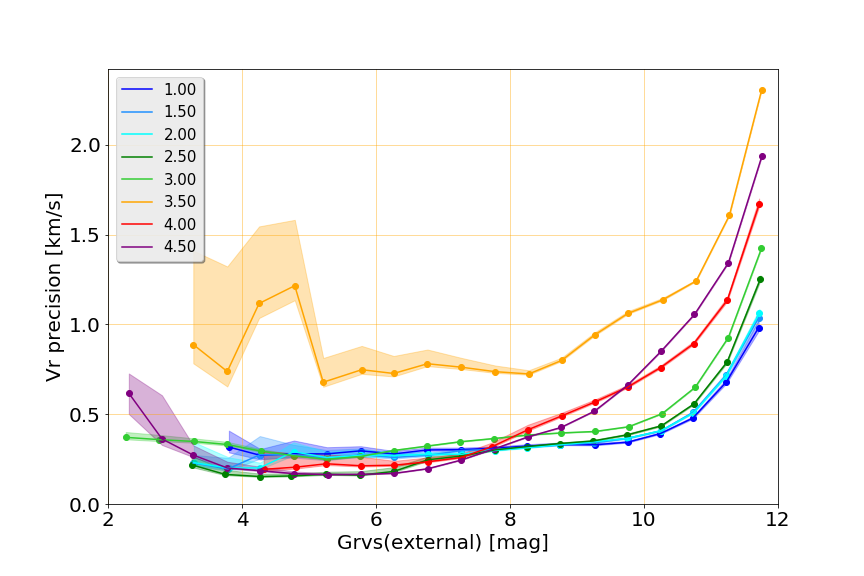}
\caption{Same as Fig.~\ref{fig:precNTr}. The curves have been calculated for different template surface gravity ranges.}
\label{fig:precTempG}
\end{figure}

As shown by Fig.~\ref{fig:precLogg} and \ref{fig:precTempG}, overall, at the faint end, the radial velocity precision improves as the, stellar and template, surface gravities decrease. The effect is more clearly visible in the full sample, which benefits from a much larger statistics and therefore more precise estimates of the precision. Stars processed with a template with surface gravity $\log g = 3.5$ show lower performance. This is the consequence of a larger mismatch between the templates and the observed RVS spectra. In the reduced library used by the software module {\it determineAP}, there is a single synthetic spectra with \tempTeff{}~$= 6000$ and a single one with 6500~K, both having a surface gravity of 3.5. The restricted library also contains 2 solar metallicity templates with an effective temperature of $5500$~K and surface gravities of 3.5 and 4.5. As discussed in Sect.~\ref{sec:teff}, stars processed with templates selected in the restricted library of 28 synthetic spectra, usually suffer from a larger template mismatch (and therefore lower radial velocity precision) than the stars with stellar parameters contained in the ground-based compilation.

\subsubsection{Precision versus metallicity}
\begin{figure}[h!]
\centering
\includegraphics[width=0.5\textwidth]{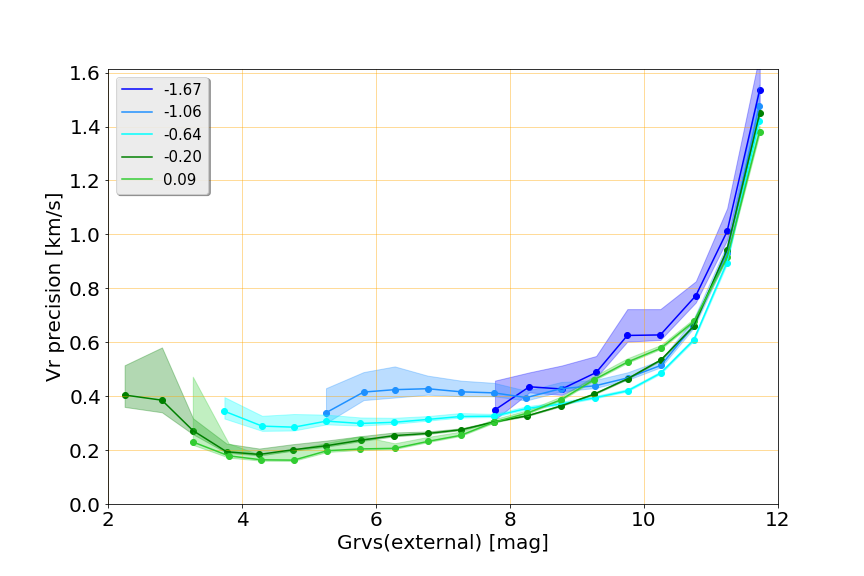}
\caption{Radial velocity precision as a function of \extgrvs{} magnitude, estimated using the full dataset. The curves have been calculated for different metallicity ranges. The mean values of the metallicity intervals are given in the caption. The lower and upper 1-$\sigma$ uncertainties on the measures of the precision are represented as shaded areas.}
\label{fig:precFeH}
\end{figure}

\begin{figure}[h!]
\centering
\includegraphics[width=0.5\textwidth]{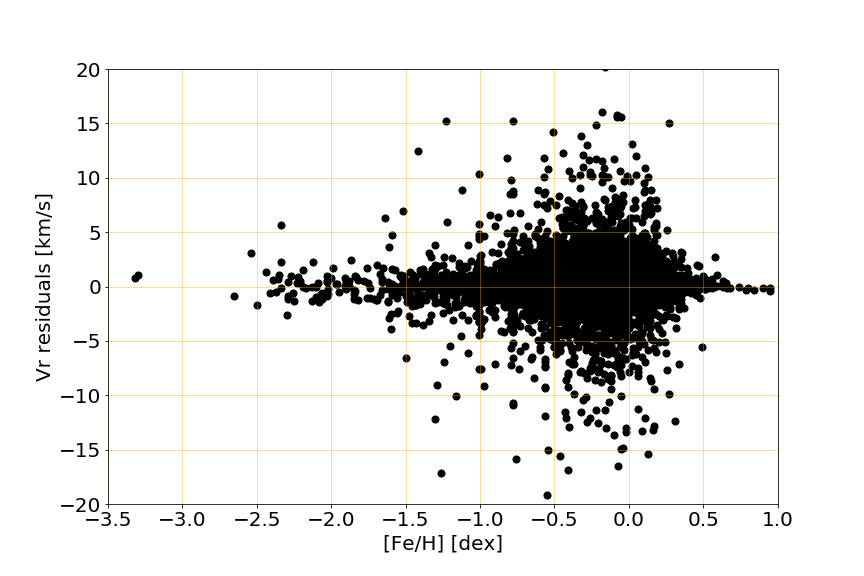}
\caption{Radial velocity residuals versus metallicity for the GB validation stars.}
\label{fig:residFeH}
\end{figure}

\begin{figure}[h!]
\centering
\includegraphics[width=0.5\textwidth]{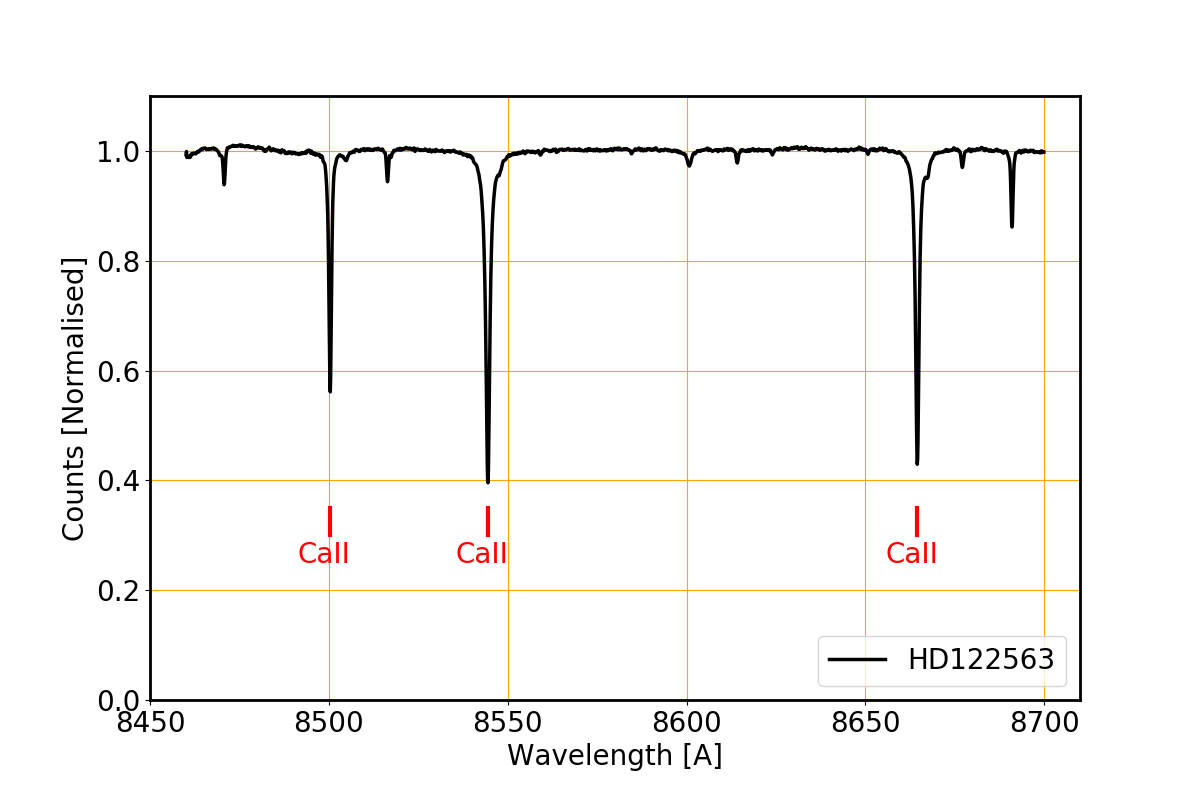}
\caption{RVS spectrum of the metal-poor star HD~122563, of metallicity $\FeH = -2.64$~dex \citep{Jofre2014}. At this metallicity, the ionised calcium triplet is still well visible.}
\label{fig:hd122563}
\end{figure}

Figure~\ref{fig:precFeH} presents the radial velocity precision as a function of the external \extgrvs{} magnitude and metallicity (the different curves), estimated using the full dataset. As expected, as the metallicity increases, the metallic lines get stronger and the radial velocity precision improves. Yet, over the metallicity range probed in Fig.~\ref{fig:precFeH}, i.e. $\sim [-2.0, 0.5]$~dex, the precision is only weakly sensitive to metallicity (e.g. compared to the sensitivity to temperature). The limited number of very metal-poor stars, with known metallicities in our compilation of ground-based catalogues, prevents to assess a reliable precision below $[Fe/H] = -2$~dex. Figure~\ref{fig:residFeH} shows the individual radial velocity residuals of the GB validation stars. Stars more metal-poor than $-2$~dex mostly show (absolute) residuals smaller than 4~$\kms$. Figure~\ref{fig:hd122563} shows the RVS spectrum of the metal-poor star HD~122563: $\Teff = 4608$~K, $\logg = 1.61$, $\FeH = -2.64$~dex \citep{Jofre2014}. Even at this very low metallicity, the ionised calcium lines are still well visible and allow to derive precise radial velocities.\\

\begin{figure}[h!]
\centering
\includegraphics[width=0.5\textwidth]{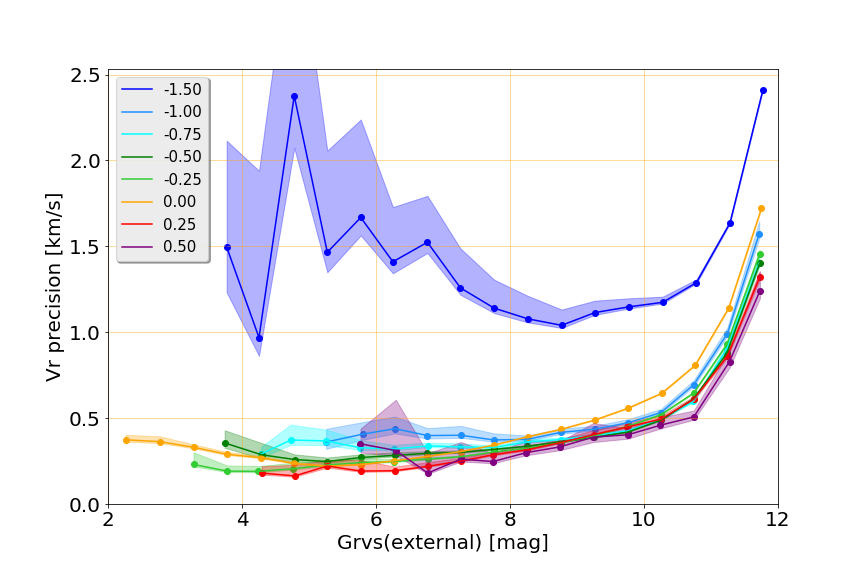}
\caption{Same as Fig.~\ref{fig:precFeH}. The curves have been calculated for different template metallicity ranges.}
\label{fig:precTempF}
\end{figure}

Figure~\ref{fig:precTempF} presents the radial velocity precision as a function of the external \extgrvs{} magnitude and metallicity of the template (the different curves), estimated with the full dataset. The reduced library of synthetic spectra used by the software module {\it determineAP} has two metallicities: $-1.5$~dex and solar. They suffer from a stronger template mismatch, than the other templates selected using the parameters from the compilation of ground-based catalogues and therefore show lower performance.

\subsection{Template selection in future \gaia{} releases}
In \gdrtwo{}, the \spectraProcessed{} spectra to process, the available processing power and the tight processing schedule have imposed a limit of 28 spectra to the library of templates used by the software module {\it determineAP}. In the subsequent \gaia{} releases, the elaborated analysis of Bp, Rp and RVS spectra (for the brightest stars) should provide precise atmospheric parameters for most \emph{RVS} stars \citep{BailerJones2013, RecioBlanco2016}, reducing the mismatch between templates and observed spectra and improving the radial velocity performances.

\subsubsection{Precision versus sky coordinates}
\begin{figure}[h!]
\centering
\includegraphics[width=0.5\textwidth]{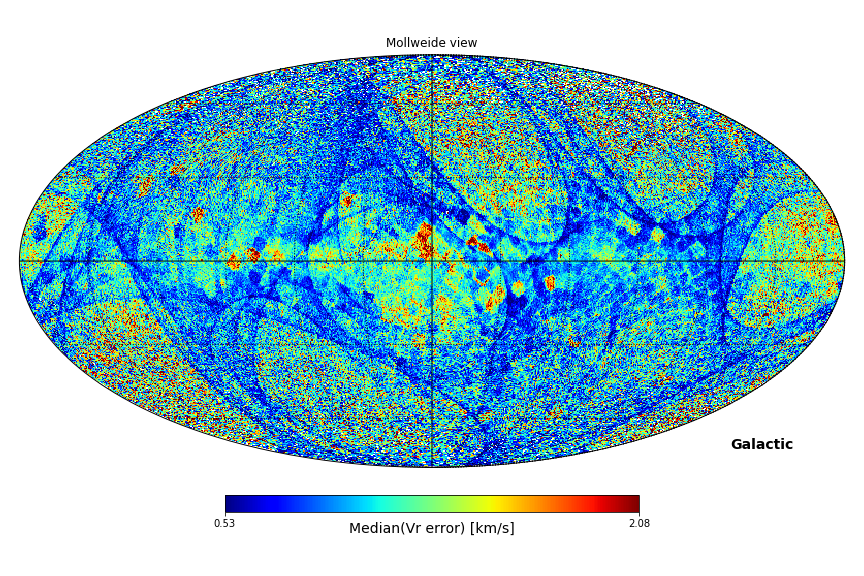}
\caption{Map of the radial velocity precision in galactic coordinates. The Galactic Centre is in the middle of the figure and the galactic longitudes increase to the left. The pixel size is 0.2 square degree (healpix level 7).}
\label{fig:precSky}
\end{figure}

Figure~\ref{fig:precSky} presents the sky map, in galactic coordinates, of the radial velocity precision estimated with the full datatset, for pixels of 0.2 square degree. The 2$^{nd}$ and 98$^{th}$ percentiles of the distribution are 0.53 and 2.08~$\kms$, while the minimum and maximum are 0.18 and 9.54~$\kms$. Comparison with Figure~\ref{fig:transit} shows that the best precisions are obtained in the area repeatedly scanned by the satellite and where the number of transits are high.

\subsubsection{Open clusters}
\begin{figure}[h!]
\centering
\includegraphics[width=0.5\textwidth]{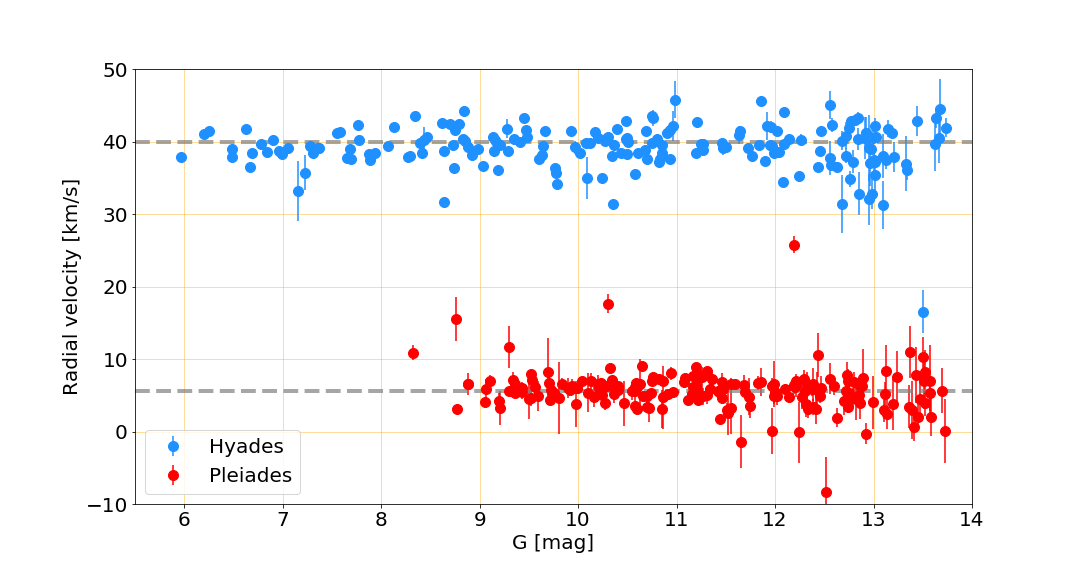}
\caption{The figure shows \gaia{} radial velocities for the astrometric members of the Hyades and Pleiades provided by \cite{DR2-DPACP-31} as a result of their membership analysis of nearby open clusters. There are nearly 200 stars in each of these two clusters having a Gaia radial velocity. The grey lines indicate the weighted averages of the radial velocity resulting from the analysis performed by \cite{DR2-DPACP-31}. The dispersion around the mean is $\sim$2~$\kms$.}
\label{fig:hyades}
\end{figure}

Open clusters are essential targets to evaluate both the consistency of radial velocities among members and the \emph{RVS} zero point by comparison to the literature. The Hyades and Pleaides are particularly well suited for that purpose because they have nearly 200 members each having a \gdrtwo{} radial velocity, and they have also been well studied with high-resolution spectroscopy. The radial velocity distribution of the astrometric members \citep{DR2-DPACP-31} as a function of $G$ magnitude (Fig.~\ref{fig:hyades}) shows a high consistency of the radial velocities on a wide magnitude range, with only the dispersion increasing at faint magnitude. The mean radial velocities of the Hyades and Pleiades are respectively $39.9\pm0.05$ and $5.55\pm0.10 \kms$, with a standard deviation of $\sim 2 \kms$, according to \cite{DR2-DPACP-31}, to be compared to $39.29\pm0.25$ and $5.94\pm0.08  \kms$ reported by \cite{Mermilliod2009}.

\begin{figure}[h!]
\centering
\includegraphics[width=0.5\textwidth]{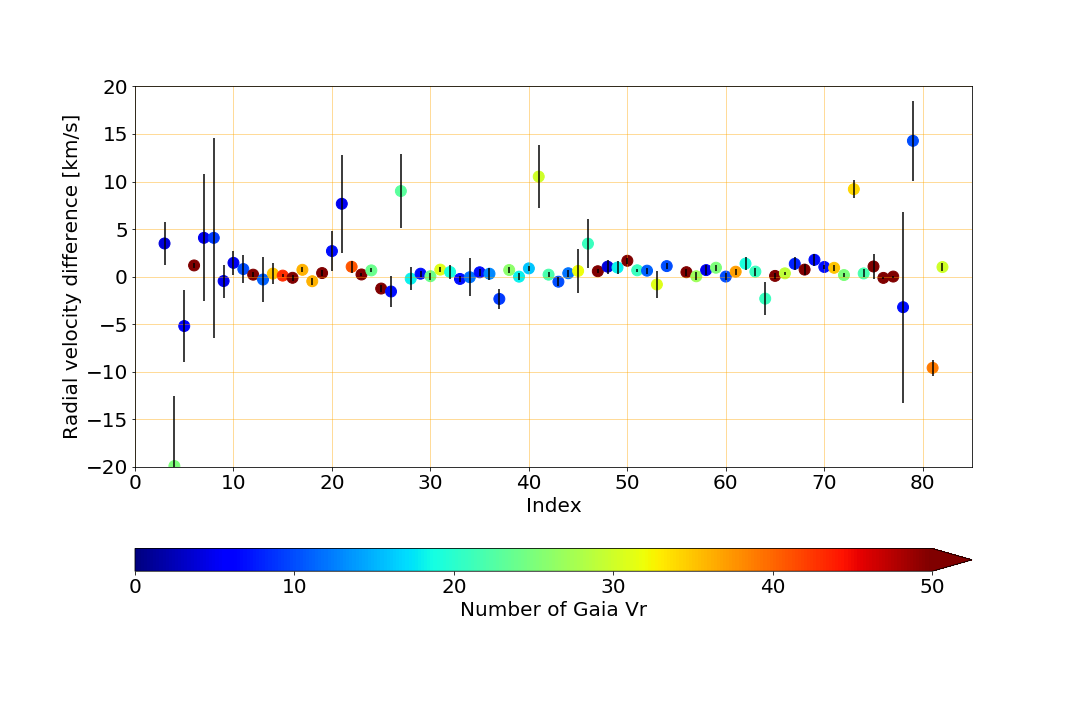}
\caption{Difference of mean radial velocities between \gaia{} and the literature \citep{Dias2002} for 82 open clusters. The error bars correspond to the quadratic sum of \gaia{} and literature uncertainties. The abscissa is an arbitrary number of the clusters from 1 to 82. The colour code highlights the number of \gaia{} members per cluster, which varies from 4 to 191.}
\label{fig:oc}
\end{figure}

Other open clusters can be used for comparison to the literature, based on astrometric membership established by \cite{DR2-DPACP-31} and \cite{DR2-DPACP-39}. The weighted mean radial velocity has been determined for open clusters having at least four member stars with a good radial velocity measurement (i.e. uncertainty smaller than 5 $\kms$) after rejection of outliers deviating by more than 10~$\kms$ from the median cluster velocity. Comparison is made with radial velocities compiled in the catalogue of \cite{Dias2002} updated in 2016. The difference of radial velocity is shown in Fig.~\ref{fig:oc} for 82 clusters in common. There are 12 open clusters with a radial velocity in disagreement by more than 5~$\kms$. The most striking difference, $\sim 80 \kms$, is for Alessi10 for which the literature value is based on only one star. Among the 12 deviating open clusters, only 3 have literature values based on at least 4 stars. For the other clusters the agreement of the \gaia{} radial velocities and that of the literature is good with a mean difference of 0.5~$\kms$ and a standard deviation of 1.2~$\kms$.

\section{Conclusion}
\gdrtwo{} contains median radial velocities for \starPublished{} stars brighter than \grvs{}~$= 12$~mag and with effective temperatures in the range $[3550, 6900]$~K. These stars offer a full sky coverage and a completeness with respect to the full second \gaia{} data release of 77.2\%, for stars with $G \leq 12.5$~mag. The accuracy of the radial velocities has been assessed by comparison to several ground-based catalogues. The medians of the radial velocity residuals vary from one catalogue to another, but do not exceed a few 100s~$\ms$. In addition, \gdrtwo{} radial velocities present a positive trend with magnitude, starting around \grvs{}~$= 9$~mag and raising up to $\sim 500\ \ms$ at \grvs{}~$= 11.75$~mag. The radial velocity precision has been assessed both by comparison with ground-based catalogues and using the distribution of \gaia{} radial velocity uncertainties. For bright stars with \grvs{} in $[4, 8]$~mag, the radial velocity precision is in the range 200-350~$\ms$. At the faint end, \grvs{}~$= 11.75$~mag, the precisions are respectively 1.4 and 3.7~$\kms$ for $\Teff = 5000$ and 6500~K.

Beyond \gdrtwo{} several catalogues are already planned\footnote{\url{https://www.cosmos.esa.int/web/gaia/release}}. Each one should include refined data treatments, new functionalities, new products and for the radial velocities a fainter processing limit. \gdrthree{} should include radial velocities for stars down to \grvs{}~$= 14$~mag, while \gdrfour{} aims at reaching the limiting magnitude of the \RVS{}: \grvs{}~$= 16.2$~mag.

\begin{acknowledgements}
This work has made use of results from the European Space Agency (ESA) space mission {\it Gaia}, the data from which were processed by the {\it Gaia} Data Processing and Analysis Consortium (DPAC). Funding for the DPAC has been provided by national institutions, in particular the institutions
participating in the {\it Gaia} Multilateral Agreement.  The {\it Gaia} mission website is \url{http://www.cosmos.esa.int/gaia}. Most of the authors are current or past members of the ESA {\it Gaia} mission team and of the {\it Gaia} DPAC and their work has been supported by the French Centre National de la Recherche Scientifique (CNRS), the Centre National d'Etudes Spatiales (CNES), the Agence Nationale de la Recherche, the R\'{e}gion Aquitaine, the Universit/'e de Bordeaux, the Utinam Institute of the Universit\'e de Franche-Comt\'e, and the Institut des Sciences de l' Univers (INSU); the Science and Technology Facilities Council and the United Kingdom Space Agency; the Belgian Federal Science Policy Office (BELSPO) through various Programme de D\'{e}veloppement d' Exp\'{e}riences Scientifiques (PRODEX) grants; the German Aerospace Agency (Deutsches Zentrum fur Luft- und Raumfahrt e.V., DLR); the Algerian Centre de Recherche 
en Astronomie, Astrophysique et G\'{e}ophysique of Bouzareah Observatory; the Swiss State Secretariat for Education, Research, and Innovation through the ESA PRODEX programme, the Mesures d' Accompagnement, the Swiss Activit\'{e}s Nationales Compl\'{e}mentaires, and the Swiss National Science Foundation; the Slovenian Research Agency (research core funding No. P1-0188). This research has made use of the SIMBAD database \citep{Wenger2000} developed and operated at CDS, Strasbourg, France.
\end{acknowledgements}

%
%

\bibliographystyle{aa} 
\bibliography{GAIA-DR2-CU6VR-DK} 

\begin{thebibliography}{34}
\expandafter\ifx\csname natexlab\endcsname\relax\def\natexlab#1{#1}\fi

\bibitem[{{Abolfathi} {et~al.}(2017){Abolfathi}, {Aguado}, {Aguilar}, {Allende
  Prieto}, {Almeida}, {Tasnim Ananna}, {Anders}, {Anderson}, {Andrews},
  {Anguiano}, \& et~al.}]{Abolfathi2017}
{Abolfathi}, B., {Aguado}, D.~S., {Aguilar}, G., {et~al.} 2017, ArXiv e-prints
  [\eprint[arXiv]{1707.09322}]

\bibitem[{{Adibekyan} {et~al.}(2012){Adibekyan}, {Sousa}, {Santos}, {Delgado
  Mena}, {Gonz{\'a}lez Hern{\'a}ndez}, {Israelian}, {Mayor}, \&
  {Khachatryan}}]{Adibekyan2012}
{Adibekyan}, V.~Z., {Sousa}, S.~G., {Santos}, N.~C., {et~al.} 2012, \aap, 545,
  A32

\bibitem[{{Anderson} \& {Francis}(2012)}]{AndersonFrancis2012}
{Anderson}, E. \& {Francis}, C. 2012, Astronomy Letters, 38, 331

\bibitem[{{Andrae} {et~al.}(2018){Andrae}, {Fouesneau}, {Creevey}, {Ordenovic},
  {Mary}, {Burlacu}, {Chaoul}, {Jean-Antoine-Piccolo}, {Kordopatis}, {Korn},
  {Lebreton}, {Panem}, {Pichon}, {Th\'evenin}, {Walmsley}, \&
  {Bailer-Jones}}]{DR2-DPACP-43}
{Andrae}, R., {Fouesneau}, M., {Creevey}, O., {et~al.} 2018, \aap\ (special
  issue for Gaia DR2)

\bibitem[{{Arenou} {et~al.}(2018){Arenou}, {Luri}, {Babusiaux}, {Fabricius},
  {Helmi}, {Muraveva}, {C. Robin}, {Spoto}, {Vallenari}, {Antoja}, {Leclerc},
  {Reyl\'e}, {Robichon}, {Shih}, {Cantat-Gaudin}, {Bossini}, {Ruiz-Dern},
  {Turon}, {Sordo}, {A. Walton}, {Blanco-Cuaresma}, {Barache}, {A. Breddels},
  {Costigan}, {Diakit\'e}, {Figueras}, {Heu}, {Jordi}, {Lallement}, {Lambert},
  {M. Marrese}, {Massari}, {Fabrizio}, {Moitinho}, {Romero-G\'omez}, {Soria},
  {Soubiran}, {Veljanoski}, {Pancino}, {Bragaglia}, {Spagna}, \&
  {Tanga}}]{DR2-DPACP-39}
{Arenou}, F., {Luri}, X., {Babusiaux}, C., {et~al.} 2018, \aap\ (special issue
  for Gaia DR2)

\bibitem[{{Bailer-Jones} {et~al.}(2013){Bailer-Jones}, {Andrae}, {Arcay},
  {Astraatmadja}, {Bellas-Velidis}, {Berihuete}, {Bijaoui}, {Carri{\'o}n},
  {Dafonte}, {Damerdji}, {Dapergolas}, {de Laverny}, {Delchambre}, {Drazinos},
  {Drimmel}, {Fr{\'e}mat}, {Fustes}, {Garc{\'{\i}}a-Torres}, {Gu{\'e}d{\'e}},
  {Heiter}, {Janotto}, {Karampelas}, {Kim}, {Knude}, {Kolka}, {Kontizas},
  {Kontizas}, {Korn}, {Lanzafame}, {Lebreton}, {Lindstr{\o}m}, {Liu},
  {Livanou}, {Lobel}, {Manteiga}, {Martayan}, {Ordenovic}, {Pichon},
  {Recio-Blanco}, {Rocca-Volmerange}, {Sarro}, {Smith}, {Sordo}, {Soubiran},
  {Surdej}, {Th{\'e}venin}, {Tsalmantza}, {Vallenari}, \&
  {Zorec}}]{BailerJones2013}
{Bailer-Jones}, C.~A.~L., {Andrae}, R., {Arcay}, B., {et~al.} 2013, \aap, 559,
  A74

\bibitem[{{Cropper} {et~al.}(2018){Cropper}, {Katz}, {Sartoretti}, {Prusti},
  {de Bruijne}, {Chassat}, {Charvet}, {Boyadijan}, {Perryman}, {Sarri}, {Gare},
  {Erdmann}, {Munari}, {Zwitter}, {Wilkinson}, {Arenou}, {Vallenari},
  {G\'{o}mez}, {Panuzzo}, {Seabroke}, {Allende Prieto}, {Benson}, {Marchal},
  {Huckle}, {Smith}, {Dolding}, {Weingrill}, {Viala}, {Blomme}, {Baker},
  {Boudreault}, {Crifo}, {Soubiran}, {Fr\'{e}mat}, {Jasniewicz}, {Guerrier},
  {Guy}, {Turon}, {Jean-Antoine}, {Th\'{e}venin}, \& {David}}]{DR2-DPACP-46}
{Cropper}, M., {Katz}, D., {Sartoretti}, P., {et~al.} 2018, \aap\ (special
  issue for Gaia DR2)

\bibitem[{{Cui} {et~al.}(2012){Cui}, {Zhao}, {Chu}, {Li}, {Li}, {Zhang}, {Su},
  {Yao}, {Wang}, {Xing}, {Li}, {Zhu}, {Wang}, {Gu}, {Luo}, {Xu}, {Zhang},
  {Liu}, {Zhang}, {Yang}, {Cao}, {Chen}, {Chen}, {Chen}, {Chen}, {Chu}, {Feng},
  {Gong}, {Hou}, {Hu}, {Hu}, {Hu}, {Jia}, {Jiang}, {Jiang}, {Jiang}, {Jin},
  {Li}, {Li}, {Li}, {Liu}, {Liu}, {Lu}, {Mao}, {Men}, {Qi}, {Qi}, {Shi},
  {Tang}, {Tao}, {Wang}, {Wang}, {Wang}, {Wang}, {Wang}, {Wang}, {Wang},
  {Wang}, {Wang}, {Wang}, {Wang}, {Wang}, {Xu}, {Xu}, {Yang}, {Yu}, {Yuan},
  {Yuan}, {Zhai}, {Zhang}, {Zhang}, {Zhang}, {Zhao}, {Zhou}, {Zhou}, {Zhu}, \&
  {Zou}}]{Cui2012}
{Cui}, X.-Q., {Zhao}, Y.-H., {Chu}, Y.-Q., {et~al.} 2012, Research in Astronomy
  and Astrophysics, 12, 1197

\bibitem[{{Dias} {et~al.}(2002){Dias}, {Alessi}, {Moitinho}, \&
  {L{\'e}pine}}]{Dias2002}
{Dias}, W.~S., {Alessi}, B.~S., {Moitinho}, A., \& {L{\'e}pine}, J.~R.~D. 2002,
  \aap, 389, 871

\bibitem[{{Evans} {et~al.}(2018){Evans}, {Riello}, {De Angeli}, {M. Carrasco},
  {Montegriffo}, {Fabricius}, {Jordi}, {Palaversa}, {Diener}, {Busso},
  {Cacciari}, \& {van Leeuwen}}]{DR2-DPACP-40}
{Evans}, D., {Riello}, M., {De Angeli}, F., {et~al.} 2018, \aap\ (special issue
  for Gaia DR2)

\bibitem[{{Gaia~Collaboration}
  {et~al.}(2018{\natexlab{a}}){Gaia~Collaboration}, {Babusiaux}, {van Leeuwen},
  {Barstow}, {Jordi}, {Vallenari}, {Bossini}, {Bressan}, {Cantat-Gaudin}, {van
  Leeuwen}, {Brown}, {Prusti}, {de Bruijne}, {Bailer-Jones}, {Biermann},
  {Evans}, {Eyer}, {Jansen}, {Klioner}, {Lammers}, {Lindegren}, {Luri},
  {Mignard}, {Panem}, {Pourbaix}, {Randich}, {Sartoretti}, {Siddiqui},
  {Soubiran}, {Walton}, {Zurbach}, \& {Zwitter}}]{DR2-DPACP-31}
{Gaia~Collaboration}, {Babusiaux}, C., {van Leeuwen}, F., {et~al.}
  2018{\natexlab{a}}, \aap\ (special issue for Gaia DR2)

\bibitem[{{Gaia~Collaboration}
  {et~al.}(2018{\natexlab{b}}){Gaia~Collaboration}, {Brown}, {Vallenari},
  {Prusti}, {de Bruijne}, {Babusiaux}, {Bailer-Jones}, {Biermann}, {Evans},
  {Eyer}, {Jansen}, {Jordi}, {Klioner}, {Lammers}, {Lindegren}, {Luri},
  {Mignard}, {Panem}, {Pourbaix}, {Randich}, {Sartoretti}, {Siddiqui},
  {Soubiran}, {van Leeuwen}, {Walton}, {Zurbach}, \& {Zwitter}}]{DR2-DPACP-36}
{Gaia~Collaboration}, {Brown}, A., {Vallenari}, A., {et~al.}
  2018{\natexlab{b}}, \aap\ (special issue for Gaia DR2)

\bibitem[{{Gaia Collaboration} {et~al.}(2016{\natexlab{a}}){Gaia
  Collaboration}, {Brown}, {Vallenari}, {Prusti}, {de Bruijne}, {Mignard},
  {Drimmel}, {Babusiaux}, {Bailer-Jones}, {Bastian}, \& et~al.}]{GaiaBrown2016}
{Gaia Collaboration}, {Brown}, A.~G.~A., {Vallenari}, A., {et~al.}
  2016{\natexlab{a}}, \aap, 595, A2

\bibitem[{{Gaia Collaboration} {et~al.}(2016{\natexlab{b}}){Gaia
  Collaboration}, {Prusti}, {de Bruijne}, {Brown}, {Vallenari}, {Babusiaux},
  {Bailer-Jones}, {Bastian}, {Biermann}, {Evans}, \& et~al.}]{GaiaPrusti2016}
{Gaia Collaboration}, {Prusti}, T., {de Bruijne}, J.~H.~J., {et~al.}
  2016{\natexlab{b}}, \aap, 595, A1

\bibitem[{{Gilmore} {et~al.}(2012){Gilmore}, {Randich}, {Asplund}, {Binney},
  {Bonifacio}, {Drew}, {Feltzing}, {Ferguson}, {Jeffries}, {Micela}, \&
  et~al.}]{Gilmore2012}
{Gilmore}, G., {Randich}, S., {Asplund}, M., {et~al.} 2012, The Messenger, 147,
  25

\bibitem[{{Houk} \& {Cowley}(1975)}]{Houk1975}
{Houk}, N. \& {Cowley}, A.~P. 1975, {University of Michigan Catalogue of
  two-dimensional spectral types for the HD stars. Volume I. Declinations -90\_
  to -53.}

\bibitem[{{Jackson} {et~al.}(2015){Jackson}, {Jeffries}, {Lewis}, {Koposov},
  {Sacco}, {Randich}, {Gilmore}, {Asplund}, {Binney}, {Bonifacio}, {Drew},
  {Feltzing}, {Ferguson}, {Micela}, {Neguerela}, {Prusti}, {Rix}, {Vallenari},
  {Alfaro}, {Allende Prieto}, {Babusiaux}, {Bensby}, {Blomme}, {Bragaglia},
  {Flaccomio}, {Francois}, {Hambly}, {Irwin}, {Korn}, {Lanzafame}, {Pancino},
  {Recio-Blanco}, {Smiljanic}, {Van Eck}, {Walton}, {Bayo}, {Bergemann},
  {Carraro}, {Costado}, {Damiani}, {Edvardsson}, {Franciosini}, {Frasca},
  {Heiter}, {Hill}, {Hourihane}, {Jofr{\'e}}, {Lardo}, {de Laverny}, {Lind},
  {Magrini}, {Marconi}, {Martayan}, {Masseron}, {Monaco}, {Morbidelli},
  {Prisinzano}, {Sbordone}, {Sousa}, {Worley}, \& {Zaggia}}]{Jackson2015}
{Jackson}, R.~J., {Jeffries}, R.~D., {Lewis}, J., {et~al.} 2015, \aap, 580, A75

\bibitem[{{Jofr{\'e}} {et~al.}(2014){Jofr{\'e}}, {Heiter}, {Soubiran},
  {Blanco-Cuaresma}, {Worley}, {Pancino}, {Cantat-Gaudin}, {Magrini},
  {Bergemann}, {Gonz{\'a}lez Hern{\'a}ndez}, {Hill}, {Lardo}, {de Laverny},
  {Lind}, {Masseron}, {Montes}, {Mucciarelli}, {Nordlander}, {Recio Blanco},
  {Sobeck}, {Sordo}, {Sousa}, {Tabernero}, {Vallenari}, \& {Van
  Eck}}]{Jofre2014}
{Jofr{\'e}}, P., {Heiter}, U., {Soubiran}, C., {et~al.} 2014, \aap, 564, A133

\bibitem[{{Kordopatis} {et~al.}(2013){Kordopatis}, {Gilmore}, {Steinmetz},
  {Boeche}, {Seabroke}, {Siebert}, {Zwitter}, {Binney}, {de Laverny},
  {Recio-Blanco}, {Williams}, {Piffl}, {Enke}, {Roeser}, {Bijaoui}, {Wyse},
  {Freeman}, {Munari}, {Carrillo}, {Anguiano}, {Burton}, {Campbell}, {Cass},
  {Fiegert}, {Hartley}, {Parker}, {Reid}, {Ritter}, {Russell}, {Stupar},
  {Watson}, {Bienaym{\'e}}, {Bland-Hawthorn}, {Gerhard}, {Gibson}, {Grebel},
  {Helmi}, {Navarro}, {Conrad}, {Famaey}, {Faure}, {Just}, {Kos}, {Matijevi{\v
  c}}, {McMillan}, {Minchev}, {Scholz}, {Sharma}, {Siviero}, {de Boer}, \& {{\v
  Z}erjal}}]{Kordopatis2013}
{Kordopatis}, G., {Gilmore}, G., {Steinmetz}, M., {et~al.} 2013, \aj, 146, 134

\bibitem[{{Kunder} {et~al.}(2017){Kunder}, {Kordopatis}, {Steinmetz},
  {Zwitter}, {McMillan}, {Casagrande}, {Enke}, {Wojno}, {Valentini},
  {Chiappini}, {Matijevi{\v c}}, {Siviero}, {de Laverny}, {Recio-Blanco},
  {Bijaoui}, {Wyse}, {Binney}, {Grebel}, {Helmi}, {Jofre}, {Antoja}, {Gilmore},
  {Siebert}, {Famaey}, {Bienaym{\'e}}, {Gibson}, {Freeman}, {Navarro},
  {Munari}, {Seabroke}, {Anguiano}, {{\v Z}erjal}, {Minchev}, {Reid},
  {Bland-Hawthorn}, {Kos}, {Sharma}, {Watson}, {Parker}, {Scholz}, {Burton},
  {Cass}, {Hartley}, {Fiegert}, {Stupar}, {Ritter}, {Hawkins}, {Gerhard},
  {Chaplin}, {Davies}, {Elsworth}, {Lund}, {Miglio}, \& {Mosser}}]{Kunder2017}
{Kunder}, A., {Kordopatis}, G., {Steinmetz}, M., {et~al.} 2017, \aj, 153, 75

\bibitem[{{Lindegren} {et~al.}(2018){Lindegren}, {Hern{\'a}ndez}, {Bombrun},
  {Klioner}, {Bastian}, {Ramos-Lerate}, {de Torres}, {Steidelm\"{u}ller},
  {Stephenson}, {Hobbs}, {Lammers}, {Biermann}, {Geyer}, {Hilger}, {Michalik},
  {Stampa}, {McMillan}, {Casta{\~n}eda}, {Clotet}, {Comoretto}, {Davidson },
  {Fabricius }, {Gracia}, {Hambly }, {Hutton}, {Mora }, {Portell }, {van
  Leeuwen}, {Abbas }, {Abreu}, {Altmann}, {Andrei }, {Anglada},
  {Balaguer-N\'u\~{n}ez}, {Barache}, {Becciani}, {Bertone}, {Bianchi},
  {Bouquillon}, {Bourda}, {Br{\"u}semeister}, {Bucciarelli}, {Busonero},
  {Buzzi}, {Cancelliere}, {Carlucci}, {Charlot}, {Cheek}, {Crosta}, {Crowley},
  {de Bruijne}, {de Felice}, {Drimmel }, {Esquej}, {Fienga}, {Fraile}, {Gai },
  {Garralda}, {Gonz{\'a}lez-Vidal }, {Guerra}, {Hauser}, {Hofmann}, {Holl},
  {Jordan}, {Lattanzi}, {Lenhardt}, {Liao}, {Licata}, {Lister}, {L{\"o}ffler},
  {Marchant}, {Martin-Fleitas}, {Messineo}, {Mignard }, {Morbidelli}, {Poggio},
  {Riva}, {Rowell}, {Salguero}, {Sarasso}, {Sciacca}, {Siddiqui}, {Smart},
  {Spagna}, {Steele}, {Taris}, {Torra}, {van Elteren}, {van Reeven}, \&
  {Vecchiato}}]{DR2-DPACP-51}
{Lindegren}, L., {Hern{\'a}ndez}, J., {Bombrun}, A., {et~al.} 2018, \aap\
  (special issue for Gaia DR2)

\bibitem[{{Majewski} {et~al.}(2017){Majewski}, {Schiavon}, {Frinchaboy},
  {Allende Prieto}, {Barkhouser}, {Bizyaev}, {Blank}, {Brunner}, {Burton},
  {Carrera}, {Chojnowski}, {Cunha}, {Epstein}, {Fitzgerald}, {Garc{\'{\i}}a
  P{\'e}rez}, {Hearty}, {Henderson}, {Holtzman}, {Johnson}, {Lam}, {Lawler},
  {Maseman}, {M{\'e}sz{\'a}ros}, {Nelson}, {Nguyen}, {Nidever}, {Pinsonneault},
  {Shetrone}, {Smee}, {Smith}, {Stolberg}, {Skrutskie}, {Walker}, {Wilson},
  {Zasowski}, {Anders}, {Basu}, {Beland}, {Blanton}, {Bovy}, {Brownstein},
  {Carlberg}, {Chaplin}, {Chiappini}, {Eisenstein}, {Elsworth}, {Feuillet},
  {Fleming}, {Galbraith-Frew}, {Garc{\'{\i}}a}, {Garc{\'{\i}}a-Hern{\'a}ndez},
  {Gillespie}, {Girardi}, {Gunn}, {Hasselquist}, {Hayden}, {Hekker}, {Ivans},
  {Kinemuchi}, {Klaene}, {Mahadevan}, {Mathur}, {Mosser}, {Muna}, {Munn},
  {Nichol}, {O'Connell}, {Parejko}, {Robin}, {Rocha-Pinto}, {Schultheis},
  {Serenelli}, {Shane}, {Silva Aguirre}, {Sobeck}, {Thompson}, {Troup},
  {Weinberg}, \& {Zamora}}]{Majewski2017}
{Majewski}, S.~R., {Schiavon}, R.~P., {Frinchaboy}, P.~M., {et~al.} 2017, \aj,
  154, 94

\bibitem[{{Makarov} \& {Unwin}(2015)}]{MakarovUnwin2015}
{Makarov}, V.~V. \& {Unwin}, S.~C. 2015, \mnras, 446, 2055

\bibitem[{{Martell} {et~al.}(2017){Martell}, {Sharma}, {Buder}, {Duong},
  {Schlesinger}, {Simpson}, {Lind}, {Ness}, {Marshall}, {Asplund},
  {Bland-Hawthorn}, {Casey}, {De Silva}, {Freeman}, {Kos}, {Lin}, {Zucker},
  {Zwitter}, {Anguiano}, {Bacigalupo}, {Carollo}, {Casagrande}, {Da Costa},
  {Horner}, {Huber}, {Hyde}, {Kafle}, {Lewis}, {Nataf}, {Navin}, {Stello},
  {Tinney}, {Watson}, \& {Wittenmyer}}]{Martell2017}
{Martell}, S.~L., {Sharma}, S., {Buder}, S., {et~al.} 2017, \mnras, 465, 3203

\bibitem[{{Mermilliod} {et~al.}(2009){Mermilliod}, {Mayor}, \&
  {Udry}}]{Mermilliod2009}
{Mermilliod}, J.-C., {Mayor}, M., \& {Udry}, S. 2009, \aap, 498, 949

\bibitem[{{Nordstr{\"o}m} {et~al.}(2004){Nordstr{\"o}m}, {Mayor}, {Andersen},
  {Holmberg}, {Pont}, {J{\o}rgensen}, {Olsen}, {Udry}, \&
  {Mowlavi}}]{Nordstrom2004}
{Nordstr{\"o}m}, B., {Mayor}, M., {Andersen}, J., {et~al.} 2004, \aap, 418, 989

\bibitem[{{Recio-Blanco} {et~al.}(2016){Recio-Blanco}, {de Laverny}, {Allende
  Prieto}, {Fustes}, {Manteiga}, {Arcay}, {Bijaoui}, {Dafonte}, {Ordenovic}, \&
  {Ordo{\~n}ez Blanco}}]{RecioBlanco2016}
{Recio-Blanco}, A., {de Laverny}, P., {Allende Prieto}, C., {et~al.} 2016,
  \aap, 585, A93

\bibitem[{{Sacco} {et~al.}(2014){Sacco}, {Morbidelli}, {Franciosini},
  {Maiorca}, {Randich}, {Modigliani}, {Gilmore}, {Asplund}, {Binney},
  {Bonifacio}, {Drew}, {Feltzing}, {Ferguson}, {Jeffries}, {Micela},
  {Negueruela}, {Prusti}, {Rix}, {Vallenari}, {Alfaro}, {Allende Prieto},
  {Babusiaux}, {Bensby}, {Blomme}, {Bragaglia}, {Flaccomio}, {Francois},
  {Hambly}, {Irwin}, {Koposov}, {Korn}, {Lanzafame}, {Pancino}, {Recio-Blanco},
  {Smiljanic}, {Van Eck}, {Walton}, {Bergemann}, {Costado}, {de Laverny},
  {Heiter}, {Hill}, {Hourihane}, {Jackson}, {Jofre}, {Lewis}, {Lind}, {Lardo},
  {Magrini}, {Masseron}, {Prisinzano}, \& {Worley}}]{Sacco2014}
{Sacco}, G.~G., {Morbidelli}, L., {Franciosini}, E., {et~al.} 2014, \aap, 565,
  A113

\bibitem[{{Sartoretti} {et~al.}(2018){Sartoretti}, {Katz}, {Cropper},
  {Panuzzo}, {Seabroke}, {Viala}, {Benson}, {Blomme}, {Jasniewicz},
  {Jean-Antoine}, {Huckle}, {Smith}, {Baker}, {Crifo}, {Damerdji}, {David},
  {Dolding}, {Fr\'{e}mat}, {Gosset}, {Guerrier}, {Guy}, {Haigron},
  {Jan{\ss}en}, {Marchal}, {Plum}, {Soubiran}, {Th\'{e}venin}, {Ajaj}, {Allende
  Prieto}, {Babusiaux}, {Boudreault}, {Chemin}, {Delle Luche}, {Fabre},
  {Gueguen}, {Hambly}, {Lasne}, {Meynadier}, {Pailler}, {Panem}, {Riclet},
  {Royer}, {Tauran}, {Zurbach}, {Zwitter}, {Arenou}, {Gomez}, {Lemaitre},
  {Leclerc}, {Morel}, {Munari}, {Turon}, \& {\v{Z}erjal}}]{DR2-DPACP-47}
{Sartoretti}, P., {Katz}, D., {Cropper}, M., {et~al.} 2018, \aap\ (special
  issue for Gaia DR2)

\bibitem[{{Soubiran} {et~al.}(2018){Soubiran}, {Jasniewicz}, {Chemin},
  {Zurbach}, {Brouillet}, {Panuzzo}, {Sartoretti}, {Katz}, {Le Campion},
  {Marchal}, {Hestroffer}, {Th\'evenin}, {Crifo}, {Udry}, {Cropper},
  {Seabroke}, {Viala}, {Benson}, {Blomme}, {Jean-Antoine}, {Huckle}, {Smith},
  {G.Baker}, {Damerdji}, {Dolding}, {Fr\'{e}mat}, {Gosset}, {Guerrier}, {Guy},
  {Haigron}, {Jan{\ss}en}, {Plum}, {Fabre}, {Lasne}, {Pailler}, {Panem},
  {Riclet}, {Royer}, {Tauran}, {Zwitter}, {Gueguen}, \& {Turon}}]{DR2-DPACP-48}
{Soubiran}, C., {Jasniewicz}, G., {Chemin}, L., {et~al.} 2018, \aap\ (special
  issue for Gaia DR2)

\bibitem[{{Steinmetz} {et~al.}(2006){Steinmetz}, {Zwitter}, {Siebert},
  {Watson}, {Freeman}, {Munari}, {Campbell}, {Williams}, {Seabroke}, {Wyse},
  {Parker}, {Bienaym{\'e}}, {Roeser}, {Gibson}, {Gilmore}, {Grebel}, {Helmi},
  {Navarro}, {Burton}, {Cass}, {Dawe}, {Fiegert}, {Hartley}, {Russell},
  {Saunders}, {Enke}, {Bailin}, {Binney}, {Bland-Hawthorn}, {Boeche}, {Dehnen},
  {Eisenstein}, {Evans}, {Fiorucci}, {Fulbright}, {Gerhard}, {Jauregi}, {Kelz},
  {Mijovi{\'c}}, {Minchev}, {Parmentier}, {Pe{\~n}arrubia}, {Quillen}, {Read},
  {Ruchti}, {Scholz}, {Siviero}, {Smith}, {Sordo}, {Veltz}, {Vidrih}, {von
  Berlepsch}, {Boyle}, \& {Schilbach}}]{Steinmetz2006}
{Steinmetz}, M., {Zwitter}, T., {Siebert}, A., {et~al.} 2006, \aj, 132, 1645

\bibitem[{{Wenger} {et~al.}(2000){Wenger}, {Ochsenbein}, {Egret}, {Dubois},
  {Bonnarel}, {Borde}, {Genova}, {Jasniewicz}, {Lalo{\"e}}, {Lesteven}, \&
  {Monier}}]{Wenger2000}
{Wenger}, M., {Ochsenbein}, F., {Egret}, D., {et~al.} 2000, \aaps, 143, 9

\bibitem[{{Yanny} {et~al.}(2009){Yanny}, {Rockosi}, {Newberg}, {Knapp},
  {Adelman-McCarthy}, {Alcorn}, {Allam}, {Allende Prieto}, {An}, {Anderson},
  {Anderson}, {Bailer-Jones}, {Bastian}, {Beers}, {Bell}, {Belokurov},
  {Bizyaev}, {Blythe}, {Bochanski}, {Boroski}, {Brinchmann}, {Brinkmann},
  {Brewington}, {Carey}, {Cudworth}, {Evans}, {Evans}, {Gates}, {G{\"a}nsicke},
  {Gillespie}, {Gilmore}, {Nebot Gomez-Moran}, {Grebel}, {Greenwell}, {Gunn},
  {Jordan}, {Jordan}, {Harding}, {Harris}, {Hendry}, {Holder}, {Ivans},
  {Ivezi{\v c}}, {Jester}, {Johnson}, {Kent}, {Kleinman}, {Kniazev},
  {Krzesinski}, {Kron}, {Kuropatkin}, {Lebedeva}, {Lee}, {French Leger},
  {L{\'e}pine}, {Levine}, {Lin}, {Long}, {Loomis}, {Lupton}, {Malanushenko},
  {Malanushenko}, {Margon}, {Martinez-Delgado}, {McGehee}, {Monet}, {Morrison},
  {Munn}, {Neilsen}, {Nitta}, {Norris}, {Oravetz}, {Owen}, {Padmanabhan},
  {Pan}, {Peterson}, {Pier}, {Platson}, {Re Fiorentin}, {Richards}, {Rix},
  {Schlegel}, {Schneider}, {Schreiber}, {Schwope}, {Sibley}, {Simmons},
  {Snedden}, {Allyn Smith}, {Stark}, {Stauffer}, {Steinmetz}, {Stoughton},
  {SubbaRao}, {Szalay}, {Szkody}, {Thakar}, {Sivarani}, {Tucker}, {Uomoto},
  {Vanden Berk}, {Vidrih}, {Wadadekar}, {Watters}, {Wilhelm}, {Wyse}, {Yarger},
  \& {Zucker}}]{Yanny2009}
{Yanny}, B., {Rockosi}, C., {Newberg}, H.~J., {et~al.} 2009, \aj, 137, 4377

\bibitem[{{Zhao} {et~al.}(2012){Zhao}, {Zhao}, {Chu}, {Jing}, \&
  {Deng}}]{Zhao2012}
{Zhao}, G., {Zhao}, Y.-H., {Chu}, Y.-Q., {Jing}, Y.-P., \& {Deng}, L.-C. 2012,
  Research in Astronomy and Astrophysics, 12, 723

\end{thebibliography}


\end{document}